\documentclass[journal=jacsat,manuscript=article]{achemso}
\usepackage{amsmath}
\usepackage{physics}
\usepackage[utf8]{inputenc}
\usepackage{tablefootnote}
\usepackage{xcolor}
\usepackage{soul}
\usepackage{multirow}
\usepackage{chemformula} 
\usepackage[T1]{fontenc} 

\usepackage[colorlinks=true,allcolors=blue]{hyperref}

\author{Shashwata Chattopadhyay}
\affiliation{School of Physics, Indian Institute of Science Education and Research Thiruvananthapuram, Kerala 695551, India}
\email{shash00817@iisertvm.ac.in}
\author{Soumyadip Hazra}
\affiliation{School of Physics, Indian Institute of Science Education and Research Thiruvananthapuram, Kerala 695551, India}
\alsoaffiliation{CAMRIE, Indian Institute of Science Education and Research Thiruvananthapuram, Kerala 695551, India}
\author{Sraboni Dey}
\affiliation{School of Physics, Indian Institute of Science Education and Research Thiruvananthapuram, Kerala 695551, India}
\alsoaffiliation{INL, CNRS, Lyon, 69100, France}
\author{Kritika Sharu}
\affiliation{School of Physics, Indian Institute of Science Education and Research Thiruvananthapuram, Kerala 695551, India}
\author{Renjith Nadarajan}
\affiliation{School of Physics, Indian Institute of Science Education and Research Thiruvananthapuram, Kerala 695551, India}
\author{Arijit Kayal}
\affiliation{School of Physics, Indian Institute of Science Education and Research Thiruvananthapuram, Kerala 695551, India}
\alsoaffiliation{Heyrovsky Institute, Czech Academy of Sciences, Dolejskova 2155/3, 18200 Prague 8, Czech Republic}
\author{Joy Mitra}
\email{j.mitra@iisertvm.ac.in}
\affiliation{School of Physics, Indian Institute of Science Education and Research Thiruvananthapuram, Kerala 695551, India}
\alsoaffiliation{CAMRIE, Indian Institute of Science Education and Research Thiruvananthapuram, Kerala 695551, India}

\title{Optically Coupled Noise Spectroscopy Quantifies Defect Passivation in Monolayer MoS$_2$ FETs}

\abbreviations{xyxy}
\keywords{xx, yy}

\begin{document}

\maketitle

\begin{abstract}
Transition metal dichalcogenide monolayers are promising materials for electronic and photonic applications, yet the performance of chemical vapour deposition grown samples is severely limited by native sulphur vacancies that introduce mid-gap trap states, degrade carrier mobility, and elevate electrical noise. Here we investigate octanethiol passivation of sulphur vacancies in monolayer MoS$_2$ field effect transistors, combining x-ray photoelectron spectroscopy, photoluminescence, and Raman scattering with electrical transport and optically coupled low-frequency noise spectroscopy. Thiol treatment reduces the sulphur vacancy concentration from 7\% to 4\%, reducing the free electron density. Electrically, it increases the device resistance 35-fold while restoring gate switching with an on/off ratio of $10^4$ and improving field-effect mobility from $\sim$1 to 5 cm${^2}$/Vs. Low frequency noise spectroscopy directly quantifies the defect suppression: the Hooge parameter $\alpha_H$ drops by more than two orders of magnitude after passivation. Gate-dependent noise confirms carrier mobility fluctuation as the dominant dark noise mechanism, while optical excitation drives a crossover to carrier number fluctuation dominated noise, reflecting preferential interaction of photogenerated carriers with residual vacancy states via generation-recombination trapping, a mechanistic distinction inaccessible to gate-bias measurements alone. Density functional theory calculations corroborate these findings, showing suppression of vacancy-induced mid-gap states by more than half and partial restoration of the intrinsic bandgap. These results establish optically coupled low-frequency noise spectroscopy as a sensitive, low-cost, and non-destructive tool for quantifying defect passivation in TMDC-based devices.
\end{abstract}



\section{Introduction}
Two-dimensional (2D) transition metal dichalcogenides (TMDCs), particularly $\text{MoS}_2$, have attracted intense interest as materials for next-generation electronic and optoelectronic devices owing to their atomically thin geometry, layer-dependent direct band gap, strong electrostatic gate control, and rich excitonic physics\cite{tsai2014monolayer, obando2025vertical, wang2018synthesis}. Unlike graphene, semiconducting TMDCs support field-effect transistors (FETs) with high on/off current ratios\cite{nourbakhsh2016mos2, wu2013high}, making them well suited for low-power logic, flexible electronics, photodetectors and chemical sensors. Further, their mechanical flexibility makes them highly amenable to flexible and wearable electronics \cite{Kang2025Strain, Yang2025Wearable}.  Large-area synthesis via chemical vapour deposition (CVD) has enabled scalable fabrication of $\text{MoS}_2$ devices, yet the performance of CVD-grown monolayers remains strongly limited by structural imperfections introduced during growth \cite{hong2015exploring, bahmani2020electronic}. Experimental and theoretical investigations have identified S vacancies as the most prevalent point defects in CVD-grown $\text{MoS}_2$\cite{Zhang2020, Vancso2016, Song2017, hong2015exploring}, which renders as-grown samples $n$-type. These vacancies introduce localised mid-gap states that trap charge carriers\cite{sharu2023leveraging} and decrease carrier mobility via scattering \cite{Song2017, Cho2015}. Further, they  destabilize the threshold voltage and reduce overall device reliability, particularly when a monolayer (ML) serves as the active FET channel. Chemical passivation using alkanethiol molecules \cite{Cho2015, schwarz2023thiol} offers a practical route to mitigate the detrimental effects of S vacancies, in which the S atoms from the thiol bind directly to the undercoordinated Mo sites. Thiol passivation suppresses mid-gap states, reduces n-type doping, and defect mediated scattering, significantly affecting transport characteristics. 

\begin{figure}[t]
	\centering
	\includegraphics[width=10cm]{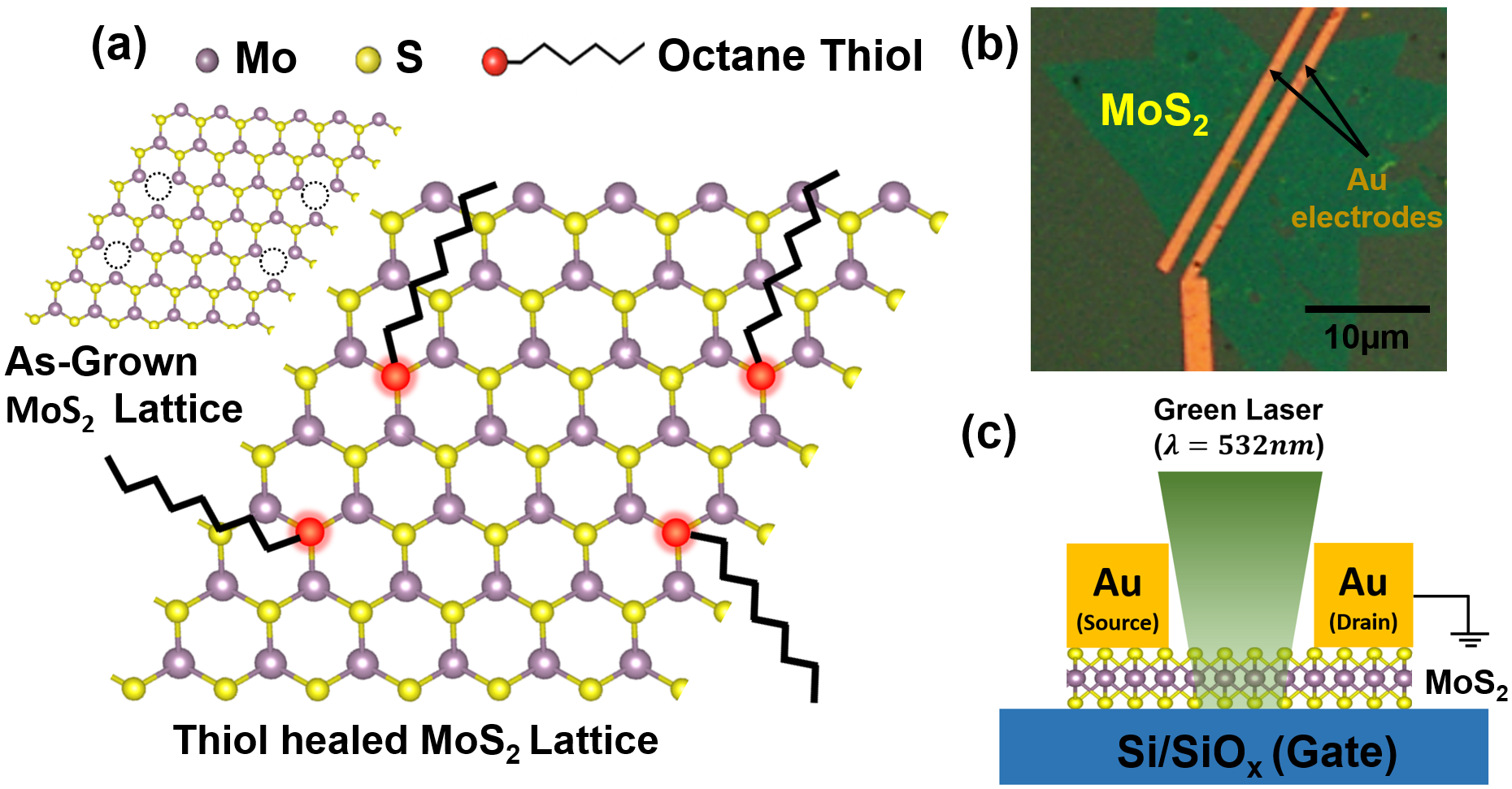}
	\caption{(a) (left) Schematic of as-grown $\text{MoS}_2$ lattice showing S vacancies, (right) S vacancy passivation by octane thiol molecules, (b) optical image of Au source drain electrodes on $\text{MoS}_2$, (c) schematic of $\text{MoS}_2$ FET device on a $\text{SiO}_x/\text{Si}$ substrate, coupled with optical excitation.}
	\label{fig:TH_MoS2}
\end{figure}

Here, we have investigated the role of octanethiol molecules in passivating surface defects on MoS$_2$ and the ensuing electrical properties. We compared the electrical properties of  FETs fabricated with CVD-grown MoS$_2$ flakes before and after thiol treatment and studied change in electrical noise in the system using low frequency  noise (LFN) spectroscopy with optical excitation. We also present Raman scattering, photoluminescence and x-ray photoelectron spectroscopy data prior to and after thiol treatment for comprehending the change in electrical response.
The top left image in fig.\ref{fig:TH_MoS2}a shows the schematic of a as-grown MoS$_2$ lattice with several S vacancies shown by dotted circles. The bottom right image then shows the envisaged passivation of defect sites by octane thiol molecules.  While Raman scattering, photoluminescence (PL), scanning probe or electron microscopy methods confirm structural and spectroscopic signatures of passivation\cite{lin2016defect}, they do not directly quantify the effect of defect healing (vacancy passivation) on charge transport, which are critical for device performance. Low-frequency noise (LFN) spectroscopy, in the spectral range $f\leq$ 1kHz, provides a direct electrical probe of transport in defect laden systems via analysing current fluctuations that typically exhibit a $1/f$ spectral dependence. In the low-frequency range, the power spectral density (PSD) of current fluctuations (S$_I(f)$) can be phenomenologically expressed as the sum of Johnson-Nyquist (thermal) noise and flicker noise described by either the Hooge formula\cite{hooge1981experimental} or the McWhorter model\cite{mcwhorter19551}. While both the latter models yield a 1/$f$ dependence of PSD, they describe two distinct mechanisms of LFN in electronic devices, namely, carrier mobility fluctuations (CMF) - Hooge's law and carrier number fluctuations (CNF) - McWhorter model. 
The empirical Hooge law expresses the current noise PSD as $S_I(f)=\alpha_H I_{ds}^2/Nf^\gamma$, where $\alpha_H$ is the Hooge parameter characterising the noise magnitude, I$_{ds}$ is the source-drain current, N is the total number of carriers given by N=$n_e$WL where $n_e$ is the carrier density, L and W are the length and width of the channel , $f$ is the frequency, and $\gamma$ is the exponent. Importantly, $\alpha_H$ and $\gamma$ serve as figures of merit for quantifying the quality of transport and $\gamma\simeq1$ for a wide variety of systems \cite{Weissman1988, Balandin2013, Dutta1981, Pal2016} .    
In metal-oxide FETs, CNF originate from the stochastic trapping-detrapping of charge carriers at the gate dielectric interface or at ionized defects in the channel leading to temporal variations in $n_e$, as given by the  McWhorter model\cite{mcwhorter19551, vonHaartman2007_LFN_MOSFET, sangwan2013low, wan2024low}. The model yields S$_I(f)\propto I_{ds}^2N_T/fC_G^2(V_G-V_{TH})^2$, where $N_T$ is the trap density, $C_G$ is the gate capacitance, and V$_G$ and V$_{TH}$ the gate voltage and threshold voltage, respectively. Note that the inverse square dependence to $C_G(V_G-V_{TH})$ also indicates S$_I(f)\propto 1/n_e^2$. Despite this sensitivity, LFN studies on 2D TMDC devices have predominantly focused on gate voltage ($\text{V}_g$) controlled carrier density fluctuations, with relatively few originating from intrinsic channel defects, i.e. S vacancies, and their modification via passivation or quenching \cite{sangwan2013low, das2015low, apl2023low, wan2024low} 
The effect of carrier density modulation on electrical transport in ML $\text{MoS}_2$ is investigated via electrostatic and photogating, for both as-grown and thiol-treated $\text{MoS}_2$. By combining electrical transport measurements with noise analysis before and after octane thiol treatment, we directly quantify the reduction in sulphur-vacancy-related trap density and its effect on carrier transport. Optical excitation at 532 nm laser serves as a non-contact gate, akin to a positive $\text{V}_g$, enabling systematic tuning of the carrier density and revealing how carriers interact with residual defect states in both as-grown and passivated devices. Density functional theory (DFT) calculations of the electronic density of states corroborate the experimental observations, providing a microscopic picture of how thiol healing suppresses in-gap states and recovers intrinsic band structure. Our results establish optically coupled LFNS as a sensitive, low-cost, and broadly applicable tool for quantifying defect passivation in TMDC-based devices.

\section{Materials and Methods}
Monolayer $\text{MoS}_2$ was synthesized via chemical vapor deposition (CVD)\cite{Lee2012SynthesisMoS2} and transferred to a p$^{++}$ doped Si substrate with 285 nm SiO$_2$. The field effect transistor (FET) devices were fabricated with the electrodes patterned via  electron beam lithography and photolithography, metallization (Au/Cr) using thermal evaporation, defining a two-probe source drain geometry on the MoS$_2$, as shown in Fig.\ref{fig:TH_MoS2}. Further details of $\text{MoS}_2$ growth and device fabrication are available in the Supporting Information (SI) section S1. 
Chemical passivation of S vacancies was performed by immersing the as-fabricated MoS$_2$ devices in a 100 mM solution of 1-octane thiol (CH$_3$(CH$_2$)$_7$SH) in ethanol for 72 hours at room temperature. Subsequently, the devices were rinsed with isopropanol and dried under a nitrogen flow prior to measurement.
Photoluminescence (PL) and Raman spectroscopy were recorded using a confocal setup (HORIBA Xplora Plus). The electrical transport measurements were carried out in a quasi four-probe geometry in a closed cycle cryostat (Advanced Research Systems), using a dc source meter (Keithley 2400) and power supply (PWS4305) . Photoconductivity measurements were conducted using a diode laser (SDL-532-040T) at 532 nm, with  intensity varied from 0 to 100\% (maximum power 5.4 mW), delivered into the cryostat through a sapphire window.
Electrical noise measurements were performed using a customised noise spectroscopy setup built around a liquid nitrogen cooled stage (LINKAM). All noise measurements were conducted between $\text{22}^o$ C and $\text{-50}^o$ C.
A constant current was applied across the device, and the resulting voltage was amplified by two custom-made low-noise preamplifiers with matched components and independent power supplies. The amplified signals were digitized using 2 independent channels of an analog-to-digital converter (NI 9250). The voltage time series were recorded over 18 second intervals and averaged over 100 successive acquisitions to further suppress random noise. The Fourier transform of the two signals was subsequently cross-correlated to cancel uncorrelated noise from all sources other than the device. The resulting power spectral density (PSD) of the current fluctuations (S$_I(f)$) was calculated in the frequency range 100 mHz to a few kHz.\cite{BarikPRA2024}. Density Functional Theory (DFT) calculations were performed using the Quantum Espresso software\cite{Giannozzi2017, Giannozzi2009} using an optimised Ultra-Soft Pseudopotentials (USPP) with Scalar-relativistic correction and non-linear core correction for all the atoms and the gen-eralised gradient approximation for the exchange correlation. Further details are available in SI Section S6.

\section{Results and Discussion}
\begin{figure}
	\centering
	\includegraphics[width=14 cm]{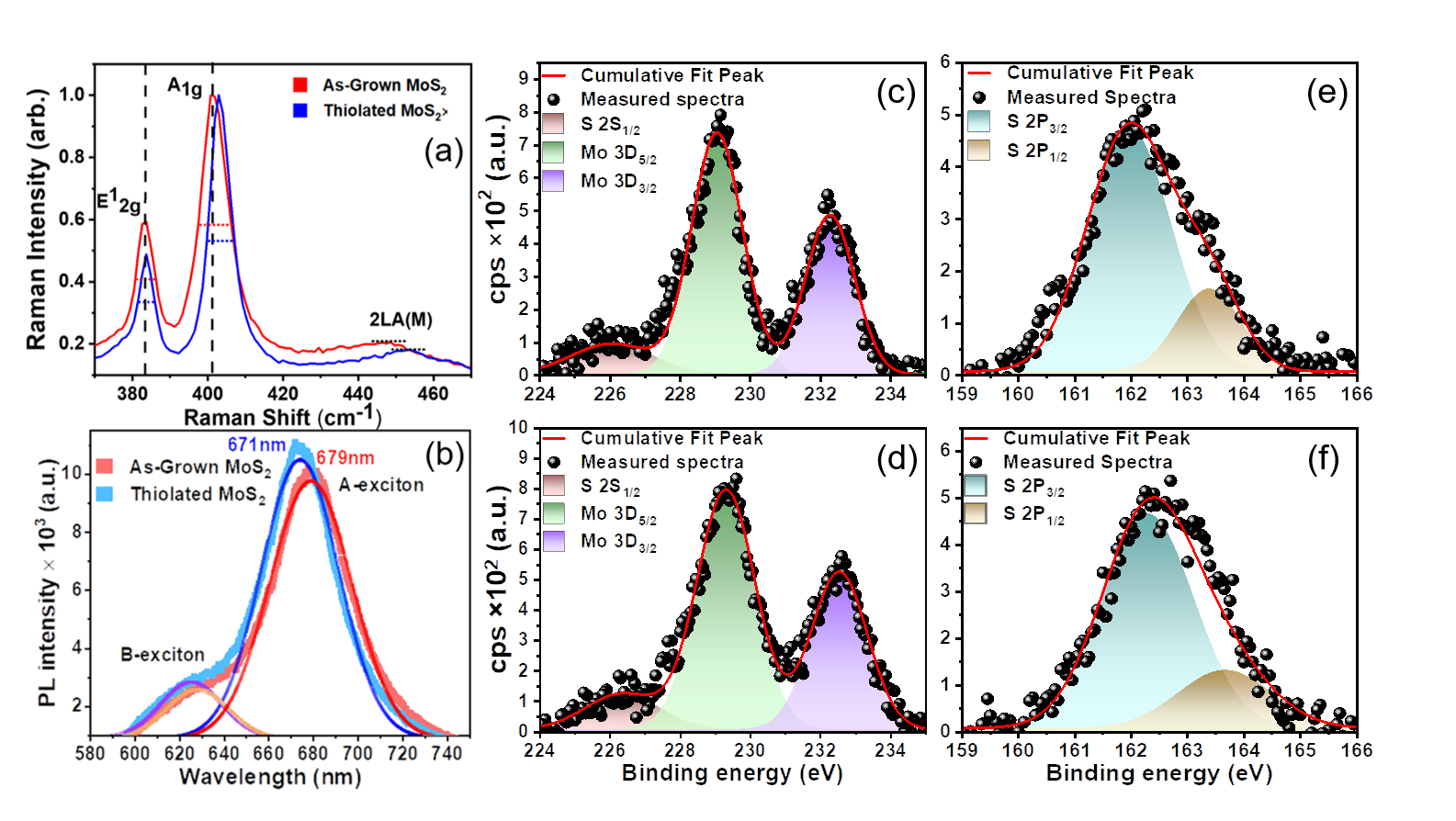}
	\caption{(a) Raman spectra of as-grown and thiol-treated monolayer MoS$_2$, showing the $E^1_{2g}$ and $A_{1g}$ modes and the $2LA(M)$ features, (b) photoluminescence  spectra before and after thiol passivation, both spectrum are deconvoluted to show excitonic contributions. XPS spectra of (c,d) Mo and (e,f) S core levels in as-grown and thiolated $\text{MoS}_2$, deconvoluted to show orbital contributions.}
	\label{fig:characteristics}
\end{figure}
Monolayers of MoS$_2$ were identified from the as-grown flakes based on the Raman spectral peaks corresponding to the A$_{1g}$ and E$^1_{2g}$ phonon modes of MoS$_2$ at 403.5 cm$^{-1}$ and 384.5 cm$^{-1}$, as shown in fig.\ref{fig:characteristics}a. The second order Raman mode 2LA(M) appears as a broad feature around 450 cm$^{-1}$ corresponding to the LA phonons at the M point of the Brillouin zone.\cite{li2012bulk,tonndorf2013photoluminescence, ionescu2017chelant,Chakraborty2012,mouri2013tunable}
Following thiol treatment on the same devices, the E$^1_{2g}$ mode energy remains unchanged, confirming that thiolation does not alter the in-plane lattice stiffness. In contrast, the A$_{1g}$ mode exhibits a blueshift of $\sim$ 3 cm$^{-1}$ and a narrowing of the FWHM by $\sim$ 3 cm$^{-1}$ (9 cm$^{-1}$ to 6 cm$^{-1}$) after thiolation, as shown in fig. \ref{fig:characteristics}a. Together, they offer direct spectroscopic signatures indicative of reduced $n_e$ (doping) in the MoS$_2$ channel, with the A$_{1g}$ mode blueshift reflecting  weakening of electron-phonon renormalisation the mode, while the linewidth reduction arises from suppressed Landau damping of the phonons. \cite{Chakraborty2012, mouri2013tunable, Chen2020DefectEngineering, Chen2025StrainDefects} Further, the ratio of the A$_{1g}$ to the 2LA(M) mode intensity, measured across four samples ({fig. S3), shows systematic increase after thiolation,  reflecting a reduction in intervalley defect scattering and confirming sulphur vacancy passivation leads to a decrease in lattice defects.

A comparison of the PL spectra acquired before and after thiol treatment shows that the A exciton emission peak blue shifts from 680 nm - 670 nm ($\Delta E\simeq$ 27 meV), as shown in \ref{fig:characteristics}(b), accompanied by a modest enhancement in the peak intensity post thiol treatment \cite{splendiani2010emerging, mouri2013tunable, birmingham2018spatially, kayal2023mobility}, which  
derives from the reduction in $n_e$ after thiol passivation. In as-grown MoS$_2$, the high background $n_e$ (S vacancy donors) favors negatively charged trions (X$^-$), which dominate PL emission redshifting the spectral weight relative to neutral exciton (X$^o$). 
Thiol passivation reduces $n_e$, suppressing the trion population and shifting spectral weight to X$^0$, which accounts for both the observed blueshift and the modest intensity enhancement.\cite{Ross2013, Mak2013} The latter arises from the increased radiative efficiency of X$^o$ over the non-radiative and trion pathways that prevail in the vacancy-rich material. Thiol passivation reduces the density of in-gap states that otherwise provide competing non-radiative $e$-$h$ recombination channels.
Thiolation studies were repeated in four samples and showed statistically comparable results as presented in SI fig. S2 and fig. S4. 

To quantify the effect of thiol passivation, the Mo:S atomic ratio was compared between samples before and after thiolation using XPS analysis. The as-grown sample shows the Mo 3d doublet (fig.\ref{fig:characteristics}c) with peaks at   $\sim$ 229 eV (3d$_{5/2}$) and $\sim$ 232.3 eV (3d$_{3/2}$) and the S 2p doublet (fig.\ref{fig:characteristics}e) at $\sim$ 162 eV (2p$_{3/2}$) and $\sim$ 163.4 eV (2p$_{1/2}$), exhibiting a spin-orbit splitting of $\sim$ 1.4 eV and intensity ratio of $\sim$ 2:1. These are consistent with the energetics of the Mo$^{4+}$ and S$^{2-}$ species in the 2H-MoS$_2$ phase. Post thiol treatment, the near uniform shift of the Mo and S doublets to higher binding energies ($\sim$ 0.3 eV), with negligible change in the spectral FWHM, is indicative of a rigid band shift rather than the emergence of altered chemical species or environments. 
This is likely to have originated from  an upward band bending induced by thiol adsorption at the MoS$_2$ surface and is consistent with previous investigations into passivation of MoS$_2$ surface using S based molecules.\cite{Cho2015,Lu2018MoS2Vacancy,Bretscher2021SulfurVacancy} 
 The average S:Mo ratio calculated for as grown samples is 1.85, which increases to 1.92 after thiol treatment and corresponds to a decrease in S vacancies from $\sim$ 7\% to $\sim$ 4\%. Details of XPS spectral analysis are available in SI section S3.

\begin{figure}[t]
	\centering
	\includegraphics[width=16cm]{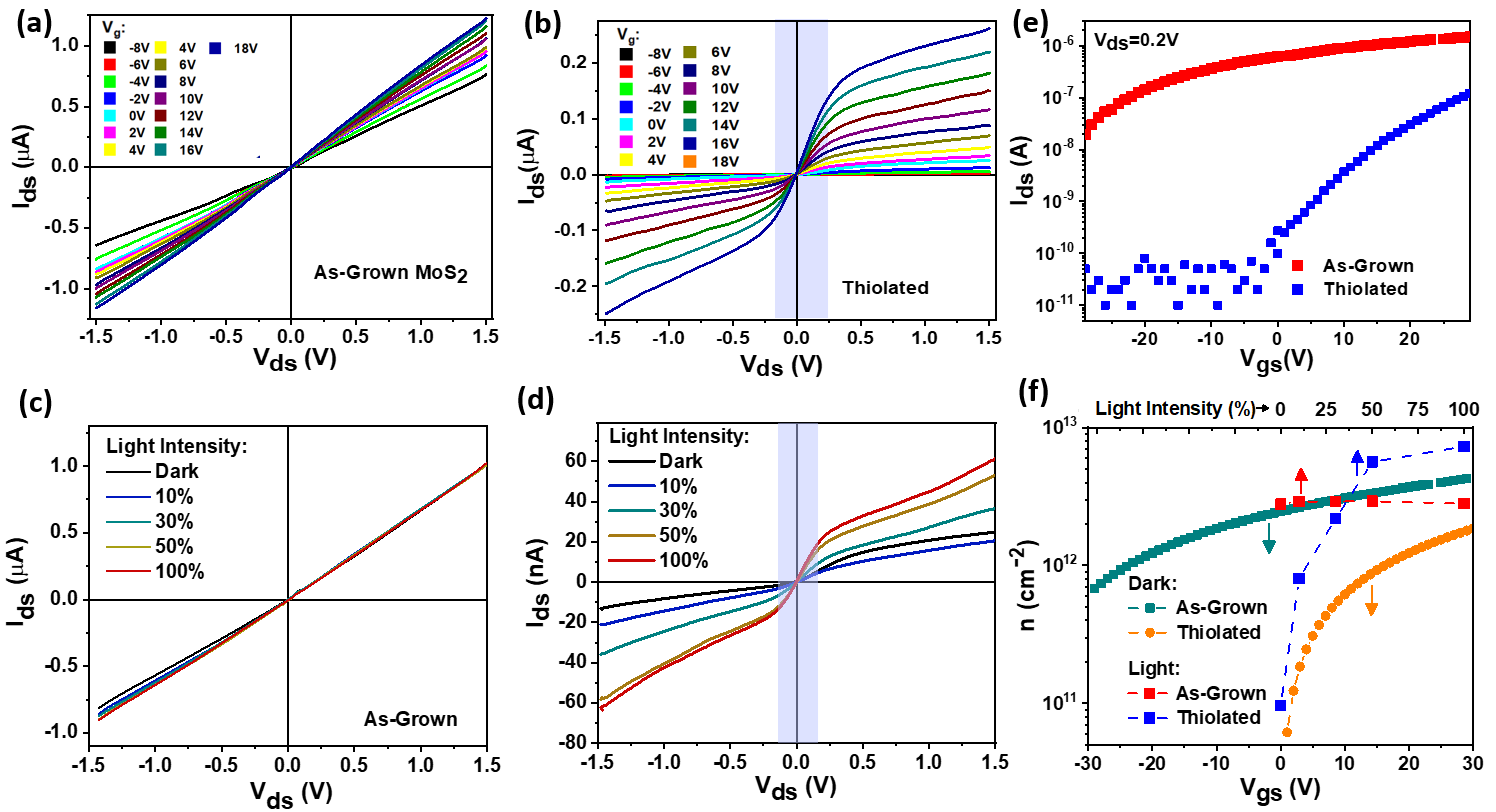}
	\caption{Device $IV$ characteristics for V$_G$ varying from -8 V to +18 V using (a) as-grown $\text{MoS}_2$ and (b) after thiol passivation (linear region highlighted in light blue). $IV$ characteristics with varying light intensity from 10\% to 100\% for (c) as-grown and (d) thiolated devices. (e) Device transfer characteristics before and after thiolation. (f) Carrier concentration ($n$) as a function of V$_G$, bottom axis) in the dark and variable light intensity (top axis) for as-grown and thiolated $\text{MoS}_2$ devices.}
	\label{fig:IV}
\end{figure}
The most prominent effect of thiolation on MoS$_2$ is observed in the electrical transport properties, shown in  fig.\ref{fig:IV}.  The dark current-voltage ($IV$) characteristics of the FET device, fabricated using an as-grown sample, show a weak non-linear behaviour along with limited gate bias (V$_G$) dependence between -8 V to +18 V (fig. \ref{fig:IV}a). The source-drain current (I$_{ds}$) non-linearity increases under positive V$_G$ yet variation of channel resistance (R$_{ch}$) and transconductance with V$_G$, depicted in SI fig.S5(d)
evidences poor gate control. This is further evidenced by the transfer characteristics (I$_{ds}$-V$_g$) in fig. \ref{fig:IV}e, where the device fails to reach 'on'`off-state' even at  V$_G$ = -30 V. The ineffective gate control is reflected in the extracted field effect mobility, $\mu_{FE} \simeq $ 1 cm$^2$/Vs (SI fig.S5(b)), which renders the device with as-grown MoS$_2$ unsuitable for FET operation. Post thiolation on the same devices, the R$_{ch}$ shows a significant increase from 1.45 M$\Omega$ to 50 M$\Omega$  (35$\times$) with the  $IV$ characteristics (fig. \ref{fig:IV}b) displaying a distinct low bias Ohmic region (|V$_{ds}| < 0.2 V$) and a 'saturation' region at higher bias. 
This large drop in conductance arises from the decrease in free electron density ($n_e$) after passivation due to quenching of the S vacancy states that act as electron donors and reduces the $n$-type nature of the material \cite{Cho2015, baik2022decreased}. Note that MoS$_2$ samples dipped in ethanol alone (without octanethiol) and washed with isopropanol were confirmed to show none of the optical or electrical changes reported above. 
The reduced conductivity ($n_e$) after thiolation denotes a reduction in $n$-type character,  independently corroborated by the Raman spectral shift of the A$_{1g}$. This thiolation induced de-doping is expected to be accompanied by a reduction of electronic E$_F$ compared to the as-grown sample, which is inconsistent with the XPS binding energy shift to higher energies, by $\sim$ 0.3 eV, as discussed earlier.   
Importantly, thiol treatment of MoS$_2$ significantly improves the degree of gate control in the FET device, as shown in \ref{fig:IV}b. The transfer characteristics (I$_{ds}$–V$_g$) shown in fig. \ref{fig:IV}e display a well-defined switch-off behavior at a threshold voltage V$_{TH} \sim 0$V \cite{kalkan2023high, kayal2023mobility}. The control afforded is reflected in the transconductance plot (SI S5(d)) that exhibits a monotonic increase with V$_G$. These features along with the I$_{ds}$ on-off ratio $\sim 10^4$ between V$_G$ = $\pm$ 30 V quantify the effective channel conductivity modulation and demonstrate that thiol functionalization substantially enhances the suitability of the device for FET applications.
Such devices also show $\mu_{FE}$ enhancement for V$_G$>0, which increases from 0.1 to 5 cm$^{2}$V$^{-1}$s$^{-1}$ between V$_G$ =$\pm$30 V SI fig.S5(b). 
The enhancement tracks V$_G$ induced modulation of channel carrier concentration ($n_e$) as shown in fig.\ref{fig:IV}f, calculated for V$_{ds}$ = 0.2 V using the relation n$_e$= C$_{ox}$(V$_G$ – V$_{TH}$)/$e$, where C$_{ox}$ is the gate oxide capacitance per unit area. Before passivation the channel $n_e$ show variation between 2 - 4 $\times 10^{12}$/cm$^2$ or V$_G$ = 0 - 30 V, which increases to 5$\times 10^{10} -  2\times10^{12}$/ cm$^2$, after thiol treatment.

A comparable change in $n_e$ is achieved under optical excitation with photon energy ($h\nu$) exceeding the bandgap energy (E$_g$) of ML MoS$_2$ (E$_g \approx 1.85$ eV). Under 532 nm illumination, the photoresponse of devices fabricated with MoS$_2$ before and after thiol treatment are very different. The as-grown sample evidences a $\sim$5\% reduction in the zero bias R$_{ch}$, under 100\% intensity (5.4mW), compared to the dark value of 1.6 M$\Omega$ (fig. \ref{fig:IV}c and SI fig.S6), with an almost constant $n_e \simeq 3 \times 10^{12}$cm$^{-2}$, independent of light intensity (fig.\ref{fig:IV}f). 
After thiol treatment, R$_{ch}$ increases to 50 M$\Omega$, which under optical excitation decreases over 75\%, as seen in fig. \ref{fig:IV}d and SI fig.S6.  In the thiolated sample, n$_e$ is estimated to increase from $6 \times 10^{10}$/cm$^2$ in the dark to $8\times 10^{12}$/cm$^2$ under 100\% intensity.
 The enhanced photoresponse is attributed to a decrease in background $n_e$ due to S vacancy passivation and the ensuing suppression of non-radiative recombination pathways, which increases carrier lifetime, thus contributing more effectively to photoconductivity.

While the above quantify the role of static defect distribution in determining the electrical properties and photoresponse of the channel, they do not capture the influence of defect dynamics on electrical transport. LFN spectroscopy provides a suitable probe to quantify the same in FET devices, varying  V$_G$ and optical excitation intensity. 
Representative PSDs, in the frequency range 100 mHz – 1 kHz, for an MoS$_2$ device, both before and after thiol treatment, are shown in \ref{fig:noise}a for various values of I$_{sd}$ with V$_G$=0 V.   
The integrated noise ($\Delta f=0.1-10$Hz) for devices with as-grown and thiol treated MoS$_2$, scale with I$_{sd}^2$ as predicted by the Hooge relation\cite{hooge1981experimental} (SI fig.S8 (b),(e) \& (h)). 
The Hooge parameter, $\alpha_H\simeq$ 10 for the as-grown sample, is independent of I$_{sd}$ (fig.\ref{fig:noise}b) and decreases more than two orders in magnitude after thiol treatment, despite the increase in R$_{ch}$. Further, for the treated sample at zero V$_G$, i.e. device in the `off' state with constant $n_e$, $\alpha_H$ shows a systematic decrease with increasing I$_{ds}$ \cite{sharma2014electrical, wan2024low} 
The frequency exponent, $\gamma\simeq 1.1$ is higher for the as-grown sample and yields a lower value (0.98 - 1.04) for the thiol treated samples.
\begin{figure}[t]
	\centering
\includegraphics[width=1.0\linewidth]{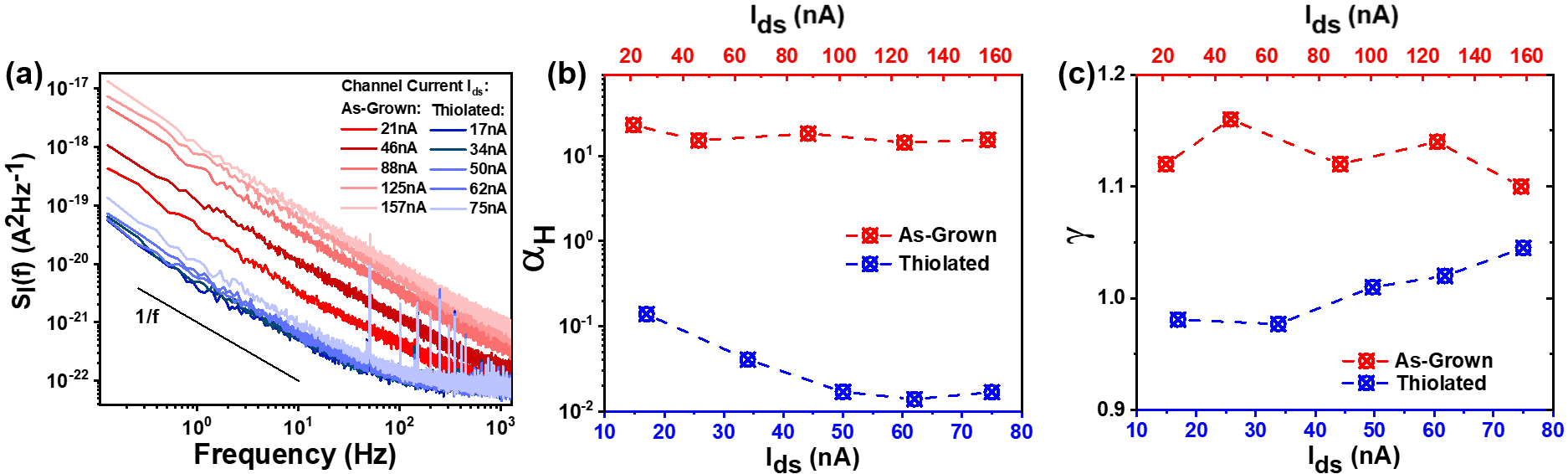}
	\caption{(a) Current noise power spectral density S$_I(f)$ measured at different I$_{ds}$, (b) Hooge parameter $\alpha$ as a function of I$_{ds}$ (c) Variation of  $\gamma$ as a function of I$_{ds}$. }
	\label{fig:noise}
\end{figure}
 
The variation of S$_I(f)$ with V$_g$ and photoexcitation intensity, both of which alter $n_e$ via different mechanisms, offers valuable insights into identifying and discriminating between noise sources i.e. CMF and CNF. The atomically thin ML channel is particularly susceptible to CMF noise due to scattering from ionized impurities (defects) like S vacancies \cite{hong2015exploring, Vancso2016}.  
LFN has been shown to decrease with increasing channel thickness \cite{kwon2014thickness, sharma2014electrical} due to decrease in scattering from surface and interfacial adsorbates that limit the mean free path length of carriers.
In addition to trapping-detrapping, CNF also arises due to photo-stimulated generation-recombination noise (GR). The former is associated with carrier capture by deep-level defect states in the MoS$_2$ bandgap and traps in the gate insulator and will be present irrespective of photo-excitation. 
Photo-stimulated GR is characterised by a range of carrier lifetimes varying from $\sim$ 1 nS (band-band transitions) to $\sim$ 1s (defect state assisted recombination), which in MoS$_2$ is dominated by defect-assisted recombination through the S vacancy states rather than band-to-band recombination, in addition to MoS$_2$/SiO$_2$ interfacial charge trapping that has been shown to prolong carrier lifetime \cite{Wang2015, DiBartolomeo2017}.
In ML MoS$_2$ transistors, LFN is likely to show a crossover between CMF and CNF as a function of electrostatic doping vs. photodoping.

\begin{figure}[t]
	\centering
	\includegraphics[width=15cm]{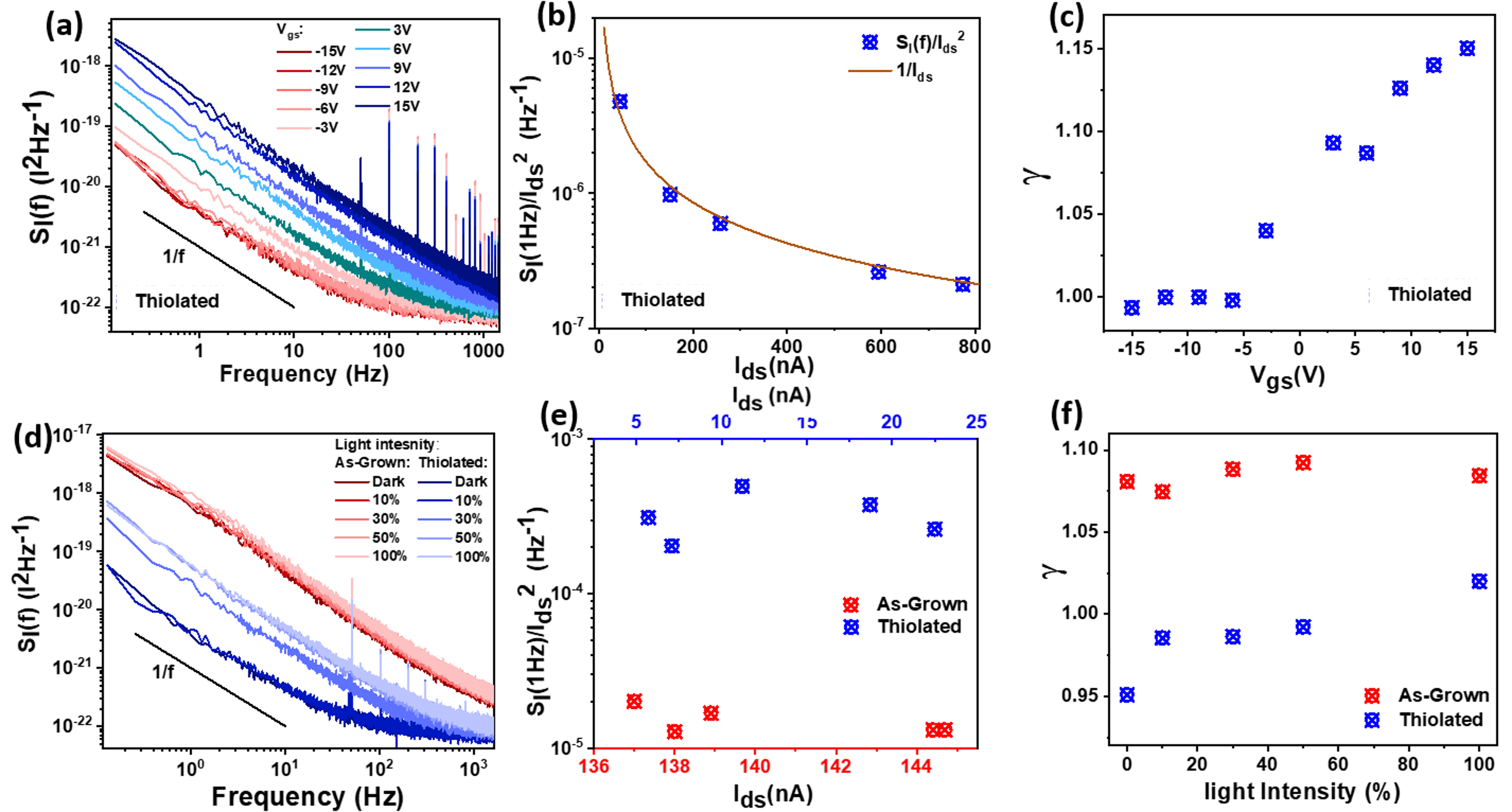}
	\caption{(a) Current noise S$_I(f)$ vs. frequency for various V$_{G}$, (b) variation of normalised S$_I(f)$/I$^2$ vs. I$_{ds}$ at $f$ = 1Hz, (c) $\gamma$ vs. V$_{G}$ for thiolated device, (d) current noise S$_I(f)$ vs. frequency under various photoexcitation intensity ($\lambda$=532nm, 100$\%$= 5.4mW) before and after thiolation, (e) variation of normalised S$_I(f)$/I$^2$ vs. I$_{ds}$ at $f$ = 1Hz, before and after thiolation, (f) $\gamma$ vs. light intensity for as-grown and thiolated device.}
	\label{fig:noise_Gate}
\end{figure}
In the dark, the noise PSD shows weak V$_G$ dependence in the ``off'' state (V$_G<0$) and increases monotonically with V$_G$ in the ``on'' state (V$_G>0$), for thiolated devices (fig.\ref{fig:noise_Gate}a). At $f$ = 1Hz, the Hooge relation predicts that the normalised noise power S$_I(1Hz)/I_{ds}^2= \alpha_H/N$, which is depicted in fig.\ref{fig:noise_Gate}b via the I$_{ds}^{-1}$ dependence of S$_I(1Hz)/I_{ds}^2$. Here, I$_{ds}\propto (V_G-V_{TH}) \propto n_e$,($n_e$=N/WL), justifying the applicability of the Hooge model and confirming CMF as the source of noise in these devices,  with an $\alpha_H\sim 0.1$ (see SI fig.S10 (a)\&(b)). The variation of the frequency exponent ($\gamma$) with V$_G$ reinforces the dominant $1/f$ noise characteristics in the ``off" state with $\gamma \simeq 1.0$, which increases to 1.15 under a positive V$_G$ = 15 V. The variation is typical of increased carrier-defect interactions expected under higher positive V$_G$ \cite{Devireddy2009}. In the McWhorter framework applicable to CNF noise, $\gamma >$1 is indicative of a non-uniform trap energy distribution or may signal the onset of correlated mobility fluctuations, consistent with a combined CNF and CMF noise mechanism in which carrier number fluctuations at trap sites are accompanied by local perturbations to carrier mobility via Coulomb scattering from the newly occupied traps. Indeed akin to the case of high positive $V_G$, under optical illumination, with high photogenerated carrier density, the noise characteristics of the devices exhibit a crossover to the CNF dominated noise as discussed below. Note that the absence of a well-defined $V_{TH}$ and the near constant $n_e$ in the device with the as-grown MoS$_2$ preclude a meaningful $V_G$ dependent Hooge model analysis of the noise PSD hence we have restricted our investigations to a thiolated device here.
 
Under optical excitation, the electrical noise PSD increases for the device fabricated from the as-grown sample (fig. \ref{fig:noise_Gate}(d) due to photogenerated carriers, which enhance trapping-detrapping dynamics at defect sites. However, thiol-treated samples show lower noise PSD compared to the as-grown $\text{MoS}_2$. Notably, the relative increase of noise as a function of light intensity is higher for passivated devices compared to the device with as-grown $\text{MoS}_2$. 
Further, unlike the gated device in the dark, under optical excitation, the noise figure S$_I(1Hz)/I_{ds}^2$ shows negligible dependence on I$_{ds}$ and thus carrier density for either the as-grown or thiolated sample devices, as shown in fig. \ref{fig:noise_Gate}e. Such weak dependence on carrier density is typically characteristic of CNF noise as described by the Mcwhorter model.  The crossover from CMF to CNF noise between gated to illuminated devices reflects the fundamentally different nature of photoexcited carriers, i.e. unlike electrostatically accumulated electrons, photogenerated electron-hole pairs interact preferentially with mid-gap S vacancy states via generation-recombination trapping dynamics, which is the hallmark of CNF noise.
The frequency exponent 
$\gamma$ remains close to unity (
0.95 - 1.02) for the thiolated device across the full range of excitation intensity (fig. \ref{fig:noise_Gate}), consistent with a McWhorter type uniform distribution of trap time constants governing the CNF process. For the as-grown device, $\gamma$ is consistently larger than that of the thiolated sample, reflecting the heterogeneous trap landscape of the vacancy-rich channel where the high density of S vacancy states with a broad distribution of trapping time constants drives the frequency exponent above unity.
In as-grown MoS$_2$, the high density of S vacancy trap states maintains a large baseline noise floor and photogenerated carriers interact with this saturated trap landscape, producing only a modest fractional increase in noise power with illumination intensity. Following passivation, the reduced trap density lowers the baseline noise substantially, making the residual active vacancy states more accessible to photogenerated carriers via generation-recombination, leading to a proportionally larger relative increase in normalised noise with $I_{ds}$ ($\propto$ illumination intensity), as shown in fig. \ref{fig:noise_Gate}e.
Unlike prior LFN studies on TMDC FETs, which have predominantly employed gate-bias modulated number fluctuations within the McWhorter framework,\cite{sangwan2013low, kwon2014thickness, ghatak2018microscopic} here, optical excitation is used to probe channel defects. Thiol treatment produces a concurrent suppression of both CMF, through diminished carrier scattering from vacancy related defect sites and CNF through a reduced density of active traps participating in generation-recombination, leading to a substantial overall reduction in LFN.

\begin{figure}[t]
    \centering
\includegraphics[width=0.8\linewidth]{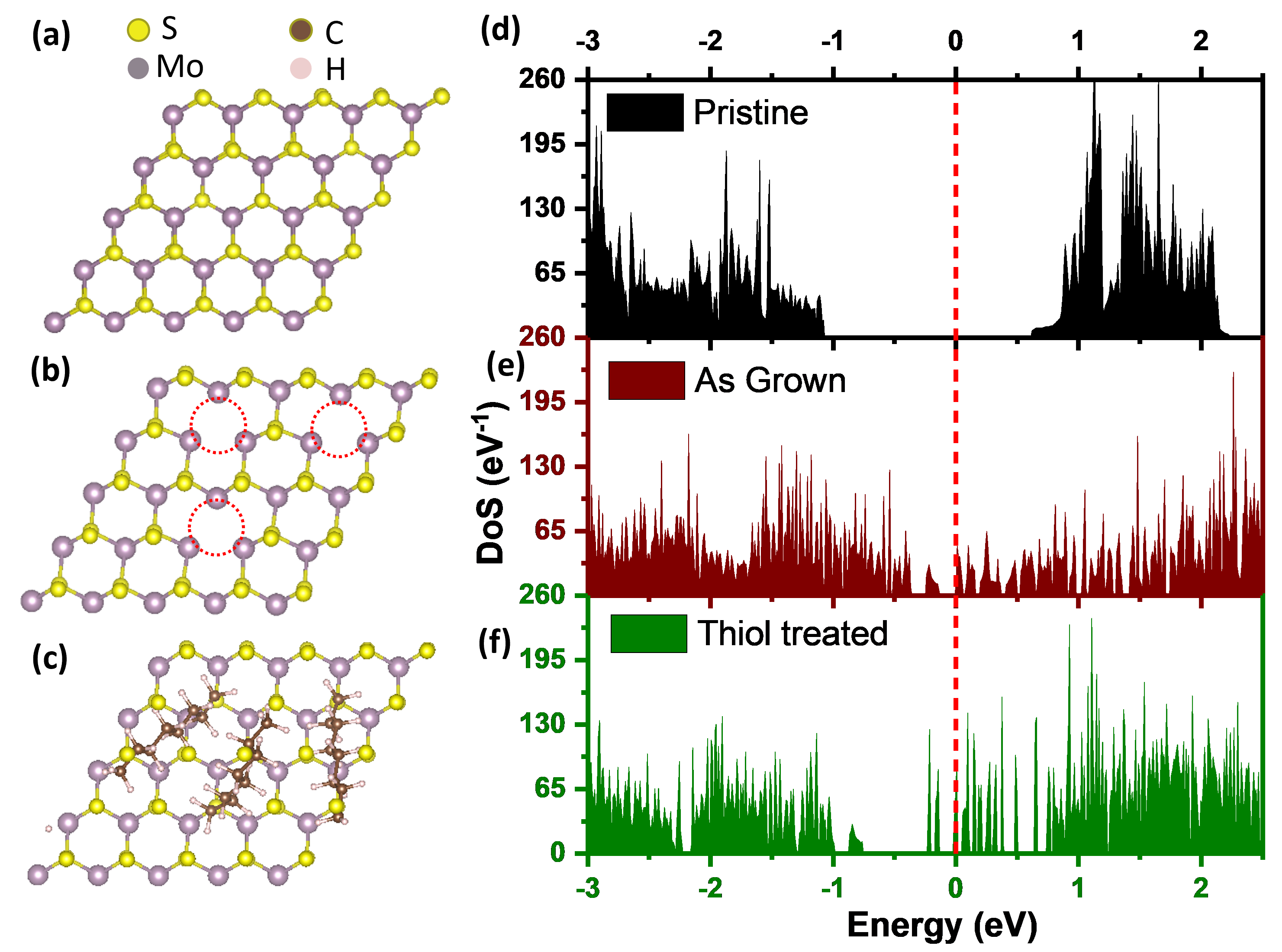}
    \caption{ 
    (a--c) Atomic models of $\mathrm{MoS}_2$ lattice (a) pristine (b) as-grown with S vacancies (dashed red circles), and (c) thiol-treated with organic molecules passivating defect sites. 
    (d--f) Corresponding density of states of $\mathrm{MoS}_2$  calculated for: (d) pristine lattice, (e) with S vacancies, showing mid-gap defect states, and (f) thiol-passivated, demonstrating partial suppression of defect states. Dashed red line shows the Fermi energy.}
    \label{fig:DFT}
\end{figure}
To provide microscopic insight into the effect of S vacancy passivation, DFT calculations were performed using the GGA-PBE functional\cite{DalCorso2014} on three configurations as shown in fig.\ref{fig:DFT}. A pristine $5\times5\times1$ MoS$_2$ supercell, a defective supercell with 6 of the 50 S atoms removed (vacancy concentration $\sim$12\%, which is higher than the experimentally determined 7\% in the as-grown samples), and a thiol-healed supercell in which the top-surface vacancy sites are passivated by octanethiol.
The calculated DOS (fig.\ref{fig:DFT}) shows that S vacancies produces a high density of mid-gap localised states that pin the $E_F$ and effectively narrows the transport gap, with states at the band edges arising from dangling bonds of undercoordinated Mo atoms acting as active trapping centres. This accounts for the high 
$n$-type conductivity, poor gate control, and elevated Hooge parameter ($\alpha_H \sim$10).  Following thiol passivation, these in-gap states are suppressed by more than 50\% as the S atoms of the octanethiol molecules bind to the dangling Mo bonds at vacancy sites. Critically, this removal of charged donor states supports the core-level shift observed in XPS, via the quenching of the donor vacancy states that reduces the surface charge density, modifying the electrostatic surface potential resulting in the upward band bending measured as a uniform binding energy shift across both Mo 3d and S 2p doublets. The negligible change in spectral FWHM confirms that this is an electrostatic rather than a chemical effect.
The $E_g$ partially recovers to $\sim$1.6 eV after thiol treatment, which is consistent with the selective passivation of the exposed MoS$_2$ surface defects. The resulting electronic inhomogeneity, i.e. a passivated channel interfacing with unpassivated areas beneath the Au contacts likely underlies the enhanced $IV$ nonlinearity observed at higher bias in the thiolated devices (fig. 3b). 
The $E_F$ shift upon passivation reflects depinning from sub-gap donor states, consistent with the $V_{TH}$ shift from -40 V to 0 V and the recovered on/off ratio of $\sim10^4$. Taken together, the DFT results provide a coherent microscopic picture in which S vacancy passivation simultaneously restores the intrinsic band structure, suppresses both CMF and CNF noise channels, and recovers the gate-tunability of the MoS$_2$ FET channel.

\section{Conclusion}
TMDC monolayers hold great promise for advancing nanoelectronic and photonic devices, yet their deployment remains constrained by structural defects, mainly S vacancies in CVD-grown MoS$_2$ that degrade carrier mobility, destabilise threshold voltage, and elevate electrical noise. We have demonstrated that optically coupled low-frequency electrical noise spectroscopy, combined with octanethiol passivation, provides a sensitive and quantitative route to characterising defect-related transport degradation and its remediation in monolayer MoS$_2$ FETs.
Thiol treatment produces a correlated set of changes across multiple probes. XPS confirms a reduction in S vacancy concentration from 7\% to 4\%, with a $\sim$0.3 eV shift commensurate with upward surface band bending. The A exciton emission peak blueshifts by $\sim$ 27 meV in PL, reflecting partial suppression of vacancy-induced sub-gap states, while Raman spectroscopy confirms lattice integrity and provides independent spectroscopic evidence of both reduced electron density and defect concentration. Electrically, the device resistance increases 35-fold, the threshold voltage shifts from $-40 V to \sim$0 V, and the field-effect mobility improves from $\sim$1 to 5 $cm^2V^{-1}s^{-1}$, changes commensurate with quenching of S defect states that sustained the high background $n$-type carrier density in the as-grown material. Optical excitation at 532 nm increases carrier density across nearly two orders of magnitude in the thiolated device, compared to the negligible change observed in the as-grown sample, demonstrating effective optical gating enabled by suppression of non-radiative recombination at vacancy sites.
LFN spectroscopy reveals the microscopic basis of these observations. The Hooge parameter 
$\alpha_H$ decreases by more than two orders of magnitude after thiolation, from $\sim$10 to $\sim$0.1, quantifying the reduction in defect-mediated scattering. In the dark, gate-dependent noise of the thiolated device confirms carrier mobility fluctuation as the dominant mechanism, with the normalised noise power $S_I/I_{ds}^2 \propto I_{ds}^{-1}$ and 
$\gamma \simeq$1.0 in the off-state, which increases toward 1.15 under positive gate bias signalling the onset of correlated mobility fluctuations at higher carrier densities. Under optical excitation, both devices exhibit a crossover to carrier number fluctuation dominated noise reflecting preferential interaction of photogenerated carriers with S vacancy states via generation-recombination trapping, distinct from the Coulomb scattering that governs electrostatically induced carriers. The concurrent suppression of both CMF and CNF after passivation accounts for the substantial overall reduction in LFN. DFT calculations provide a consistent microscopic picture: S vacancies introduce mid-gap states that pin the Fermi level and sustain elevated trapping dynamics, while thiol passivation suppresses these states by more than 50\%, partially restoring the intrinsic band structure and accounting for the improved gate control, reduced 
$n$-type doping, and lower noise floor observed experimentally.
These results establish optically coupled LFN spectroscopy as a sensitive, low-cost, and non-destructive tool for quantifying defect passivation in TMDC devices, with direct transferability to other chalcogenide systems, offering a general framework for defect characterisation and device optimisation in atomically thin semiconductor technology.

\section*{Acknowledgements}
Authors thank Prof. M. M. Shaijumon
(IISER Thiruvananthapuram) for the use of CVD growth facilities. Authors acknowledge financial support from ANRF, Government of India (CRG/2023/006878), MOE-STARS (STARS-2/2023-1012), SPARC (No. 3086), and IISER Thiruvananthapuram for computing time on the Padmanabha cluster. SD acknowledges a PhD fellowship from DST INSPIRE. RN acknowledges University Grants Commission, Govt. of India, for a
PhD fellowship. 

\section*{Author Contributions}
SC, AK and JM conceived the project. RN synthesized the MoS$_2$ samples. SC and JM wrote the manuscript with inputs from all authors.
\section*{Supporting information}

\renewcommand{\thefigure}{S\arabic{figure}}
\renewcommand{\thetable}{S\arabic{table}}
\renewcommand{\theequation}{S\arabic{equation}}
\setcounter{figure}{0}
\section{S1: Sample preparation and characterization (AFM/TEM)}
The $\text{MoS}_2$ monolayers were synthesized by chemical vapour deposition
(CVD) technique with $\text{MoO}_3$ and sulphur as the precursors. $\text{MoO}_3$
powder was dissolved in ethanol and then spin-coated onto a $\text{SiO}_2$/Si wafer. The spin-coated substrate was then placed inside a quartz tube to be heated inside the furnace. Approximately 500
mg of sulfur powder was placed at the upstream end of the tube. The
growth was then carried out at 850°C under a continuous argon
flow of 100 sccm. 
The fabrication of the field effect transistor (FET) devices on CVD grown
$\text{MoS}_2$ monolayers were conducted on $\text{SiO}_2$/Si substrates. At first, all the substrates were primarily cleaned with acetone, isopropyl alcohol (IPA)
and de-ionised (DI) water followed by cleaning in a mixture of solvents that
consisted of $\text{H}_2\text{O}_2$ +HCL+DI water (1:1:3) with constant heating at ~60$^0$C and then rinsed with DI water and dried. A 30 nm thick hydrogen
silsesquioxane (HSQ) polymer (2 wt$\%$ in MIBK) was spin-coated on the bare
$\text{SiO}_2$/Si substrates and heated at 500$^0$C for 1 hour in order to convert the polymer to $\text{SiO}_x$ which ensured a pinhole-free surface. Next, large area square (~600 $\mu$m) Au contact pads were deposited using conventional
photolithography followed by thermal evaporation of 50nm/4nm of Au/Cr
on the patterned $\text{SiO}_x$-coated substrates. Subsequently, CVD grown $\text{MoS}_2$ flakes are transferred to the pre-patterned substrates using a PMMA-
based wet etch transfer method using 2M NaOH solution. At the end, smaller
Au contacts were created on the $\text{MoS}_2$ monolayers to make contact to the larger
Au pads using standard electron-beam lithography with a positive PMMA
resist, yielding the final FET device with L=2$\mu$m and W=4.9$\mu$m. L and W being the channel length and channel width respectively. Atomic Force Microscopy (AFM) image of the fabricated FET devices are shown in \ref{fig:AFM}(a) with Au electrodes on monolayer MoS$_2$. The AFM measurements were performed using Bruker Multimode 8 AFM in ambient conditions. Topography was measured by tapping mode, and the measurement was performed using non-coated cantilever with $f_0 \sim 325$ kHz from MikroMasch (HQ:NSC15/Al-BS). 

The HRTEM image was acquired using a FEI Tecnai G2 F30 STwin transmission electron microscope (TEM) 300 kV, after transferring a CVD‑grown monolayer of MoS$_2$ onto a carbon‑support TEM grid. \ref{fig:AFM}(b) shows TEM image of the monolayer MoS$_2$ with visible sulphur vacancy sites.
\begin{figure*}
	\centering
	\includegraphics[width=0.6\linewidth]{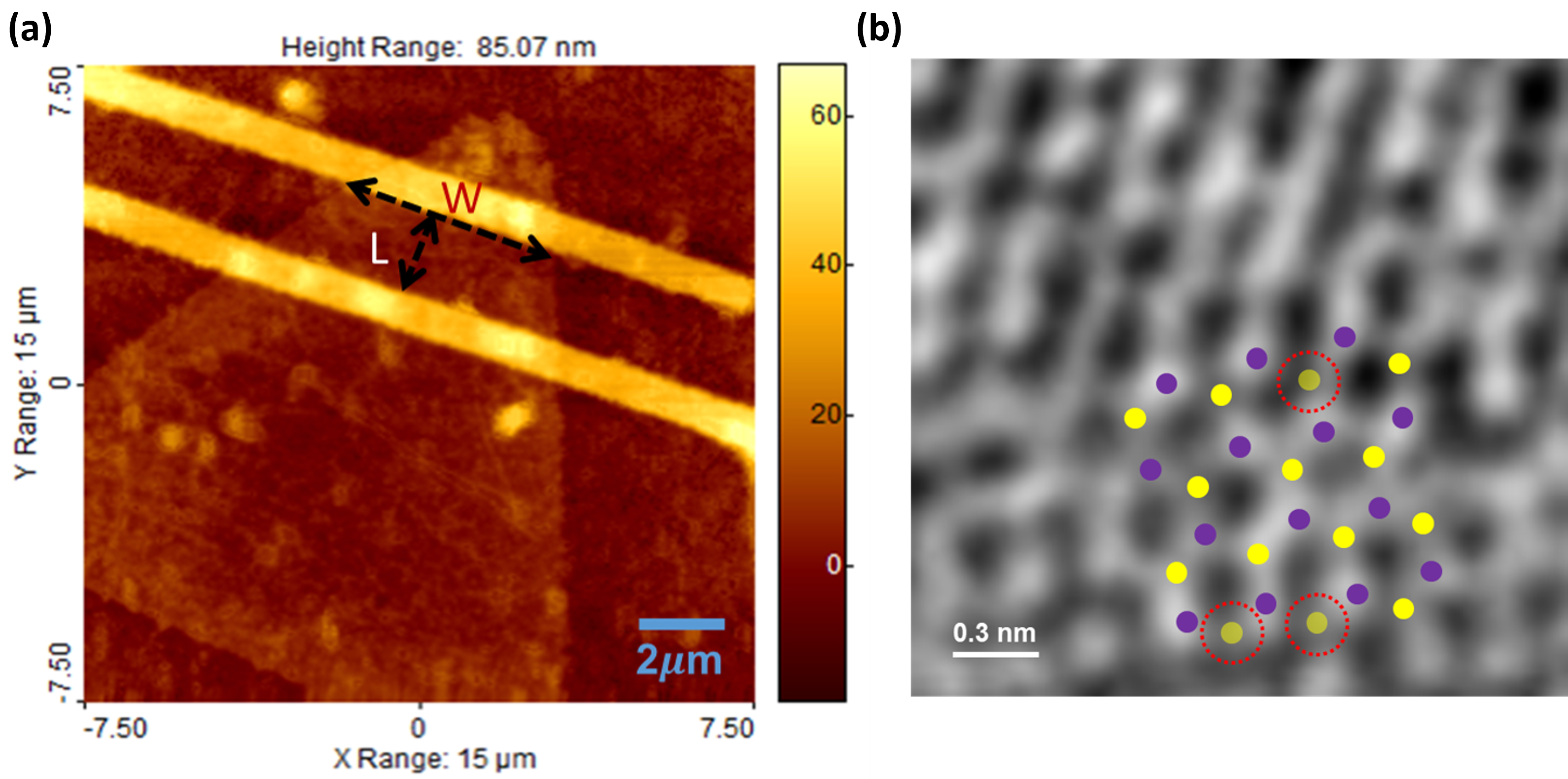}
	\caption{AFM topographic and atomic-scale defect characterization of monolayer MoS$_2$: (left) Topographic scan of a typical MoS$_2$ flake device channel with defined dimensions $L$ and $W$ (scale bar: 2~$\mu$m); (right) High-resolution TEM atomic lattice image highlighting intrinsic sulphur vacancy defect sites enclosed in red dotted circles (scale bar: 0.3~nm).}
	\label{fig:AFM}
\end{figure*}
\newpage
\subsection{S2: Photoluminescence and Raman spectroscopy}
\begin{figure}[h!]
	\centering
	\includegraphics[width=12cm]{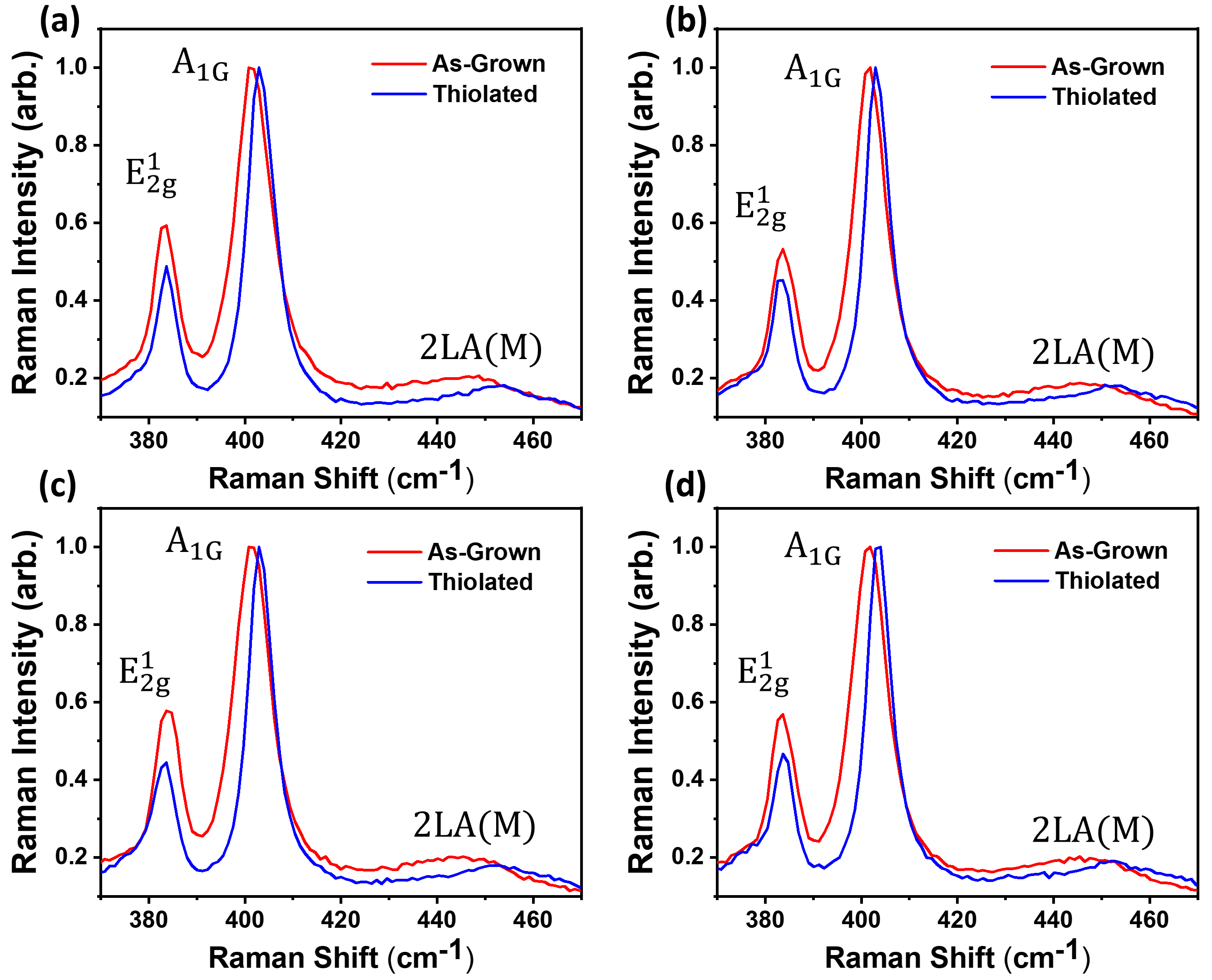}
	\caption{Raman spectra (a--d) mesasured across four devices as-grown and then thiol-treated. The spectra show a consistent blue shift of the $A_{1g}$ mode following thiol treatment, consistent with the reduction in electron density, while the $E^1_{2g}$ and $2LA(M)$ modes exhibit comparatively small changes in their spectral positions.}
	\label{fig:Raman}
\end{figure}

\begin{figure}[h!]
	\centering
	\includegraphics[width=8cm]{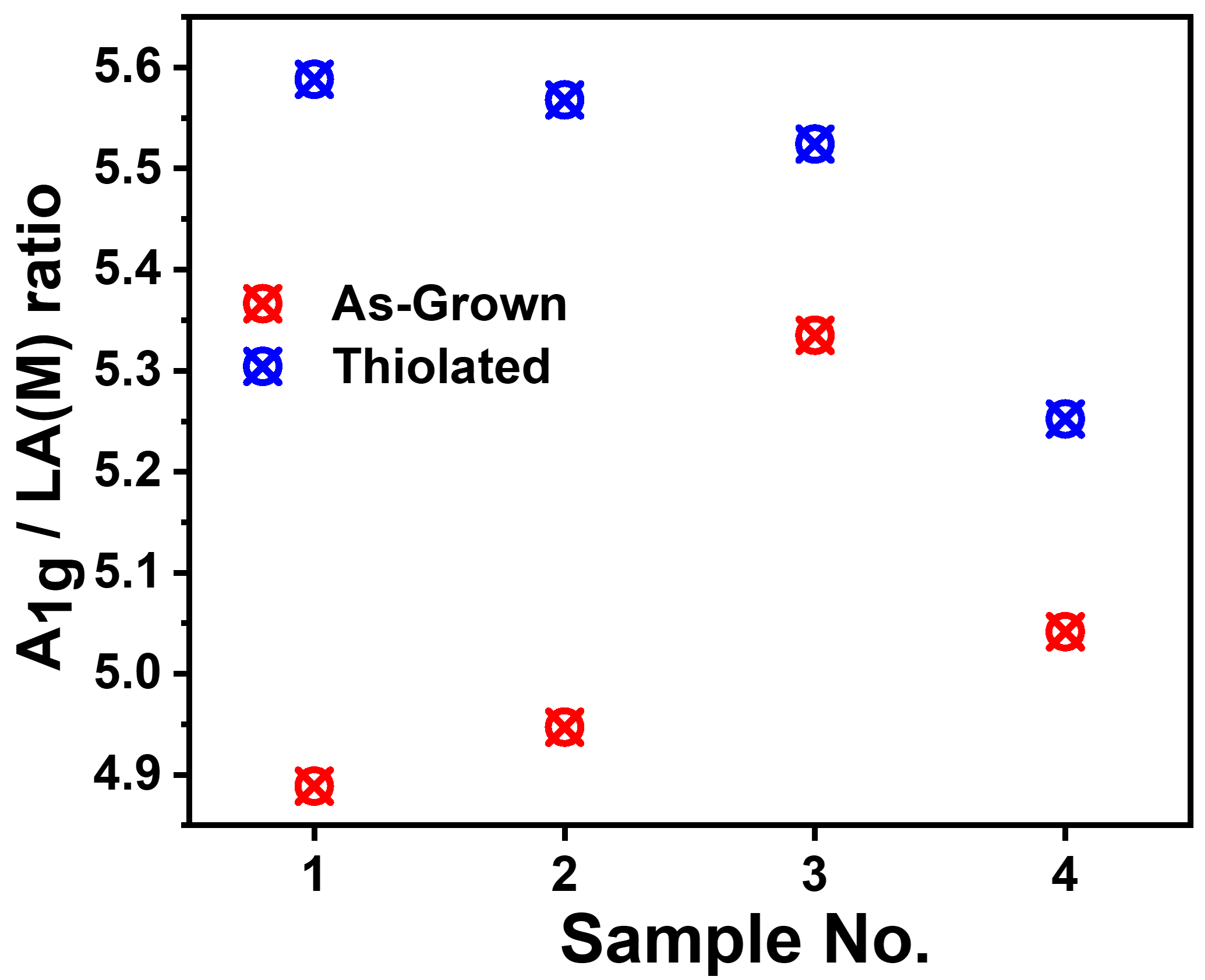}
	\caption{Ratio of the $A_{1g}$ Raman peak intensity to that of the $2LA(M)$ mode for as-grown and thiol-treated MoS$_2$ samples.}
	\label{fig:LAM}
\end{figure}

\begin{figure}[h!]
	\centering
	\includegraphics[width=15cm]{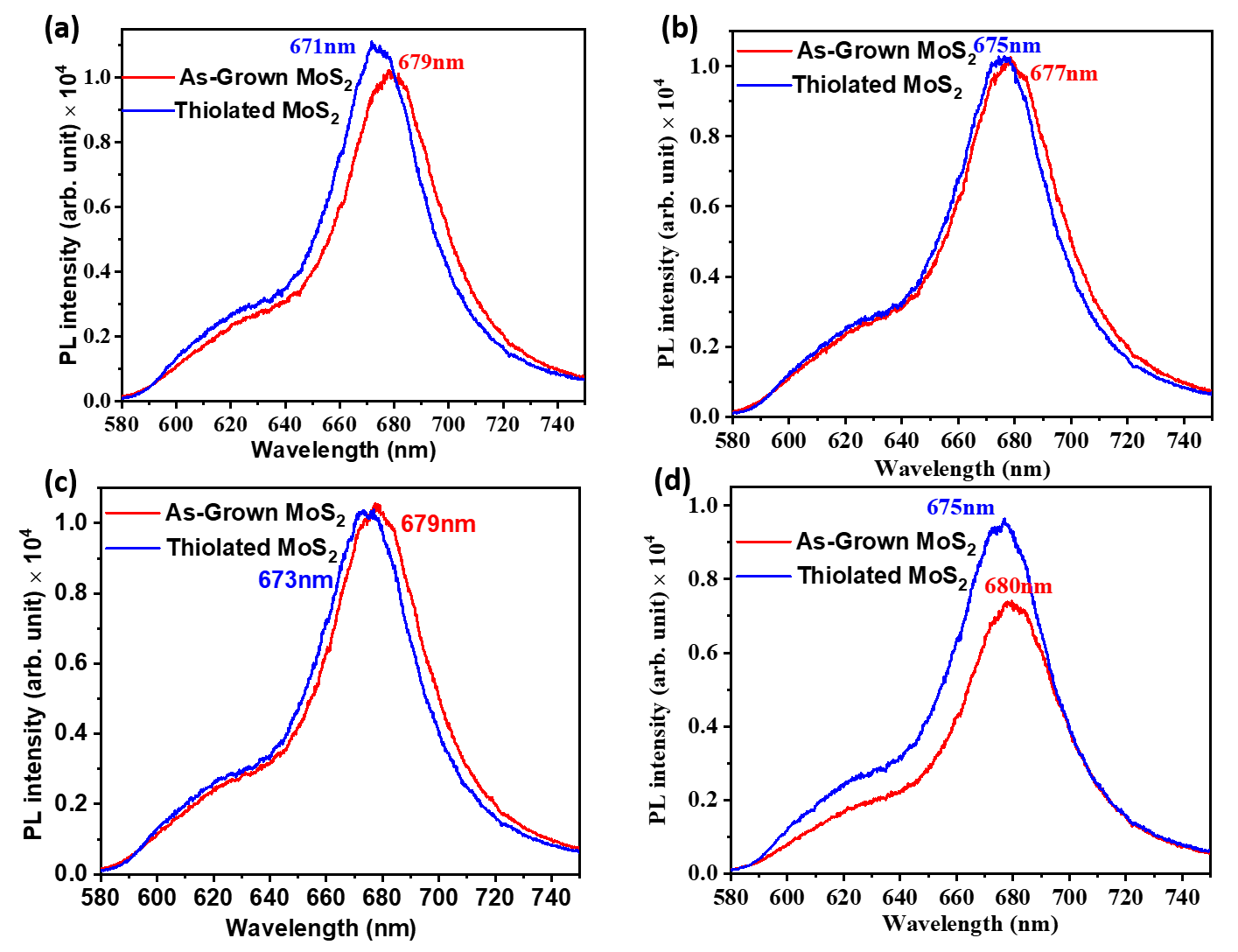}
	\caption{Normalized photoluminescence spectra (a--d) acquired from four devices of as-grown and then thiolated MoS$_2$, showing consistent blue-shifting of the emission profile and suppression of trion recombination across the flake.}
	\label{fig:PL}
\end{figure}
Photoluminescence (PL) and Raman studies were conducted using a HORIBA Xplora Plus Raman setup based on a confocal microscope. The PL and raman data  were acquired with 532 nm laser excited through a 100X objective with NA: 0.9.
Spectra for PL and Raman were recorded with TE-cooled (-$60^0$) CCD, using grating with ruling density of 600 gr/mm, and 2400 gr/mm, respectively. All the spectroscopic measurements were conducted in room temperature. Fig.\ref{fig:Raman} shows the shift in E$^1$$_{2G}$, A$_{1G}$ and the 2LA(M) mode before and after thiol treatment across 4 different samples while Fig.\ref{fig:PL} show the PL spectra across 4 different samples before and after thiol treatment. Fig.\ref{fig:LAM} shows the change in ratio of A$_{1G}$ and 2LA(M) across different samples before and after thiol treatment.

\newpage
\section{S3: XPS data}
X-ray photoelectron spectroscopy (XPS) 
measurements were performed using the Omicron Nanotech instrument with a Al K$_{\alpha}$ X-ray 
source, the CasaXPS software was used for data analysis. The carbon correction with respect to 284.8 eV was done for all the elements.
To confirm the S defect passivation by the thiol treatment, the stoichiometry calculation was performed on as grown and 1-octanethiol treated MoS$_2$, using the core level spectra of Mo 3d, S 2s and S 2p orbits from the XPS spectra. The atomic percentage ratio of the elements S and Mo is calculated using the relation 
S(at.\%)/M(at.\%) = $\langle(A_{S}/\sigma_{S}$)$\rangle$/$\langle(A_{Mo}/\sigma_{Mo}$)$\rangle$ that calculates the ratio of the average area ($A$) under the elemental peaks normalised with the photoionization cross-section ($\sigma$) at the incident photon energy of 1.5 keV\cite{scofield1973theoretical}. The peak shifts for each orbital and their corresponding photoionization cross-sections are tabulated in Table \ref{tab:xps_combined}; notably, the peak shifts between the as-grown and thiolated samples are very consistent with previous report\cite{Cho2015-xm}.

\begin{table}[h!]
\centering
\caption{XPS peak parameters for Mo and S core levels: as-grown vs.\ thiolated samples}
\label{tab:xps_combined}
\begin{tabular}{llccccc}
\hline
\textbf{Orbital} & \textbf{Sample} & \textbf{Peak (eV)} & $\boldsymbol{\Delta}$\textbf{ Peak (eV)} &  \textbf{Cross-section (Mb)\cite{scofield1973theoretical}} \\
\hline
\multirow{2}{*}{Mo $3d_{5/2}$} & As-grown  & 229.0 & \multirow{2}{*}{$+0.3$} & \multirow{2}{*}{0.7463} \\
                                & Thiolated & 229.3 &                          & \\
\hline
\multirow{2}{*}{Mo $3d_{3/2}$} & As-grown  & 232.3 & \multirow{2}{*}{$+0.2$} & \multirow{2}{*}{0.51508} \\
                                & Thiolated & 232.5 &                          &  \\
\hline
\multirow{2}{*}{S $2s_{1/2}$}  & As-grown  & 225.8 & \multirow{2}{*}{$+0.4$} & \multirow{2}{*}{0.19066} \\
                                & Thiolated & 226.2 &                          &  \\
\hline
\multirow{2}{*}{S $2p_{3/2}$}  & As-grown  & 162.0 & \multirow{2}{*}{$+0.3$} & \multirow{2}{*}{0.147} \\
                                & Thiolated & 162.3 &                          &  \\
\hline
\multirow{2}{*}{S $2p_{1/2}$}  & As-grown  & 163.4 & \multirow{2}{*}{$+0.3$} & \multirow{2}{*}{0.075216} \\
                                & Thiolated & 163.7 &                          &  \\
\hline
\end{tabular}
\end{table}

\section{S4: Electrical data acquisition}
Electrical transport measurements were performed using a Keithley 2400 Source Meter. The device was mounted inside a closed-cycle 4K cryostat (Advance Research System) under a vacuum of $10^{-3}$mbar. All electrical measurements discussed here were carried out at room temperature. The sheet conductivity $\sigma$ of the device is related to the carrier number density through:
\begin{equation}
	\sigma=ne\mu
\end{equation} 
where n is the carrier number density, e is the elementary charge, and $\mu$ is the field-effect mobility. $\sigma$ is calculated from the relation,
\begin{equation}
	\sigma=\frac{L}{W}\frac{dI_{ds}}{dV_{ds}}
\end{equation} 

The gate-dependent mobility $\mu$ is given by,
\begin{equation}
	\mu=\frac{L}{WC_gV_{ds}}\frac{dI_{ds}}{dV_{gs}}
\end{equation} 
Using this mobility, the gate-induced carrier density was calculated \cite{Cho2015},
\begin{equation}
	n=\frac{C_g}{e} (V_{gs}-V_{th})
\end{equation} 
Where $C_g$ is the gate capacitance per unit area and $V_{th}$ is the threshold voltage.To quantify the effect of illumination, the light-induced change in carrier density ${\Delta}n$ was estimated from the measured photocurrent using,
\begin{equation}
	\Delta n=\frac{\Delta I_{ds}}{q\mu(W/L)V_{ds}}
\end{equation} 
where $\Delta I_{ds} = I_{light}-I_{dark}$, q is the elementary charge, and W, L, and $V_{ds}$ have their usual meanings. This relation is derived from the linear regime current expression:
\begin{equation}
	\Delta I_{ds}= qn \mu (W/L)V_{ds}
\end{equation} 

Fig.\ref{fig:mobility}(a) shows the gate dependent channel current $I_{ds}$ trasnfer characteristics, the $V_{th}$ before and after thiol treatment corresponds to -40V and 0V respectively. Fig.\ref{fig:mobility}(b) extracted gate-dependent mobility, while Fig.\ref{fig:mobility}(c) \& (d) present the corresponding gate-dependent resistance and trans-conductance for devices before and after thiol treatment.  Fig.\ref{fig:LIV}(a) and (b) shows the current-voltage variation and change in device resistance as a function of light illumination (0-100\%, 5.4mW) before thiol treatment while Fig.\ref{fig:LIV}(c) and (d) shows the current-voltage variation and change in device resistance as a function of light intensity after thiol treatment, highlighting the combined electrostatic and photo-induced modulation of charge transport. Note, device shows negligible response before thiol treatment when transport of electrons is majorly dominated by defect induced doping.

\begin{figure}[h!]
	\centering
	\includegraphics[width=12cm]{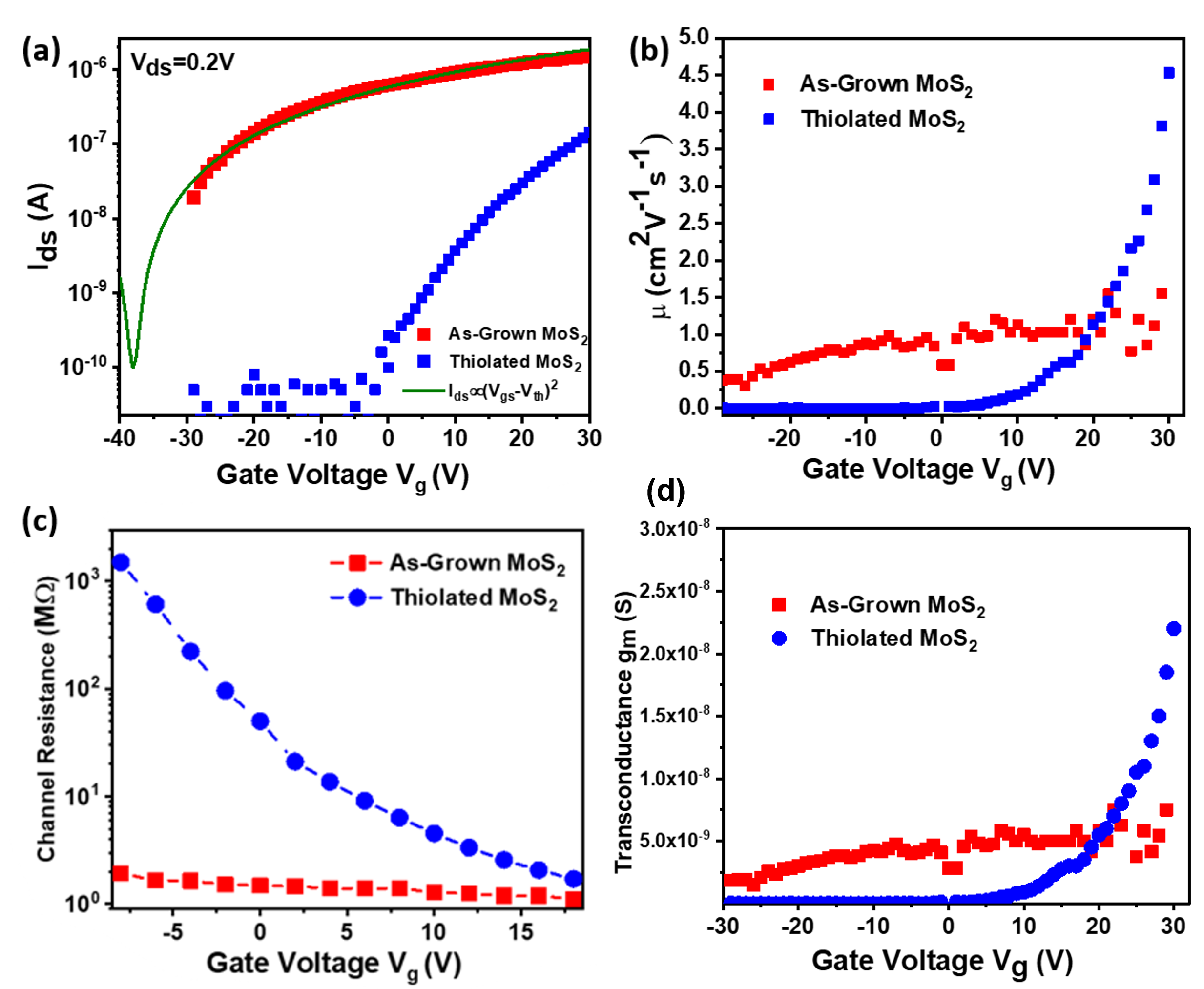}
	\caption{Field-effect mobility and channel resistance modulation: (a) Transfer characteristics ($I_{ds}$--$V_g$) at $V_{ds}=0.2$ V with a quadratic fit for the as-grown device (green solid line). (b) Extracted field-effect mobility as a function of gate voltage. (c) Gate-voltage-dependent channel resistance. (d) Corresponding transconductance as a function of gate voltage.}
	\label{fig:mobility}
\end{figure}
\begin{figure}[h!]
	\centering
	\includegraphics[width=12cm]{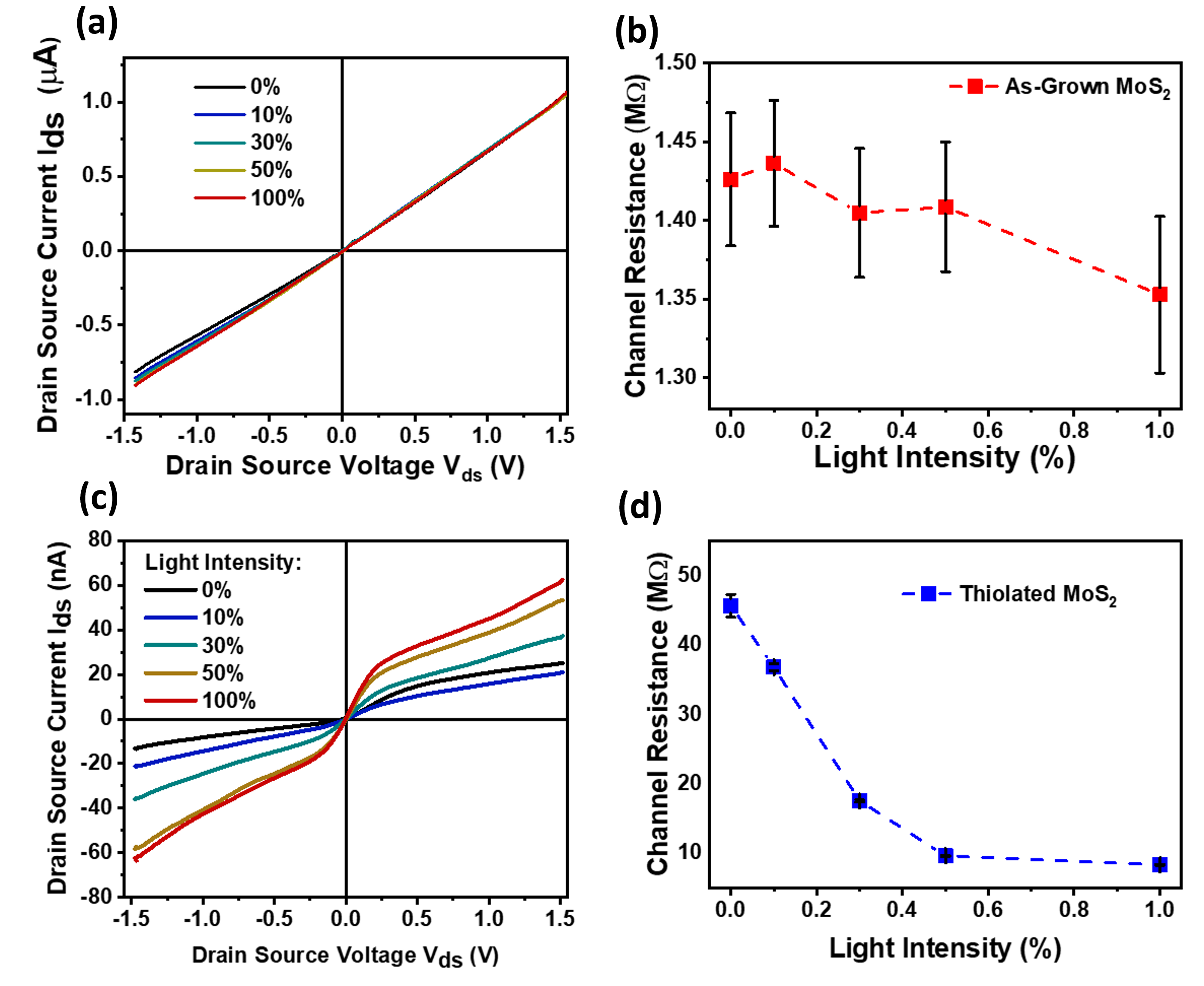}
	\caption{Photo-induced transport and channel resistance variation under 532~nm laser excitation: (a) $I_{ds}$--$V_{ds}$ curves as function of Light intensity for As-Grown devices and (b) corresponding channel resistance as a function of light intensity for the as-grown device; (c) $I_{ds}$--$V_{ds}$ curves as function of Light intensity for Thiolated device and (d) channel resistance drop as a function of light intensity for the thiolated device.}
	\label{fig:LIV}
\end{figure}
\clearpage
\section{S5: Low frequency noise}
 \begin{table*}[ht]
\centering
\caption{Summary of low-frequency noise regimes, dominant mechanisms, and key parameters 
for as-grown and thiolated MoS$_2$ FET devices under dark (electrostatic gating) 
and optical excitation conditions.}
\label{tab:noise_summary}
\renewcommand{\arraystretch}{1.5}
\begin{tabular}{p{2.8cm} p{1.8cm} p{2.2cm} p{4.5cm} p{2.2cm} p{2.0cm}}
\hline\hline
\textbf{Condition} & \textbf{Sample} & \textbf{Noise Regime} & \textbf{Dominant Mechanism} & \textbf{Key Parameter} & \textbf{Value} \\
\hline
Dark, $V_G = 0$ 
& As-grown 
& CMF 
& Mobility fluctuations from ionised S vacancy scattering 
& $\alpha_H$ 
& $\sim 10$ \\

Dark, $V_G = 0$ 
& Thiolated 
& CMF 
& Mobility fluctuations from residual vacancy scattering 
& $\alpha_H$ 
& $\sim 0.1$ \\

Dark, $V_G < 0$ (``off'') 
& Thiolated 
& CMF 
& Pure mobility fluctuations; minimal carrier-trap interaction 
& $\gamma$ 
& $\simeq 1.0$ \\

Dark, $V_G > 0$ (``on'') 
& Thiolated 
& CMF $\rightarrow$ CNF+CMF 
& Enhanced carrier-defect scattering; onset of correlated mobility fluctuations 
& $\gamma$ 
& $1.0 \rightarrow 1.15$ \newline ($V_G = 15$\,V) \\

Illuminated \newline (532\,nm, variable intensity) 
& As-grown 
& CMF $\rightarrow$ CNF 
& GR trapping at high-density S vacancy mid-gap states; modest incremental CNF due to trap saturation 
& $S_I/I_{ds}^2$ vs $I_{ds}$; $\gamma$ 
& Weak $I_{ds}$ dependence; $\gamma \simeq 1.0$--$1.05$ \\

Illuminated \newline (532\,nm, variable intensity) 
& Thiolated 
& CNF 
& GR trapping-detrapping at residual vacancy states; pronounced relative noise increase with intensity 
& $S_I/I_{ds}^2$ vs $I_{ds}$; $\gamma$ 
& Negligible $I_{ds}$ dependence; $\gamma \simeq 0.95$--$1.0$ \\

Dark, variable $V_G$ 
& As-grown 
& --- 
& Poor gate control; near-constant $n_e \simeq 2$--$4 \times 10^{12}$\,cm$^{-2}$; $V_{TH} \simeq -40$\,V 
& $\mu_{FE}$ 
& $\simeq 1$\,cm$^2$V$^{-1}$s$^{-1}$ \\

Dark, variable $V_G$ 
& Thiolated 
& --- 
& Good gate control; $n_e = 5\times10^{10}$--$2\times10^{12}$\,cm$^{-2}$; $V_{TH} \simeq 0$\,V 
& $\mu_{FE}$ 
& $0.1$--$5$\,cm$^2$V$^{-1}$s$^{-1}$ \\

\hline\hline
\end{tabular}
\end{table*}

The devices were kept in a LINKAM stage for low temperature mesurement in a vaccum pressure of $10^{-2}mBar$ and temperture varying between 22$^0$C to -50$^0$C by pumping liquid nitrogen into the LINKAM stage. The contact pad on the device was connected using copper wires and the cupper wires were coupled outside the LINKAM stage using BNC connectors.
The two preamplifiers used to amplify the noise signal from device under test (DUT) was custom designed and self-made. It uses a two stage inverted OPAMP amplifiers configuration. Each stage proving a gain of 33 and a total combined gain of the preamplifiers being 1089. 

\begin{figure*}
	\centering
	\includegraphics[width=1.0\linewidth]{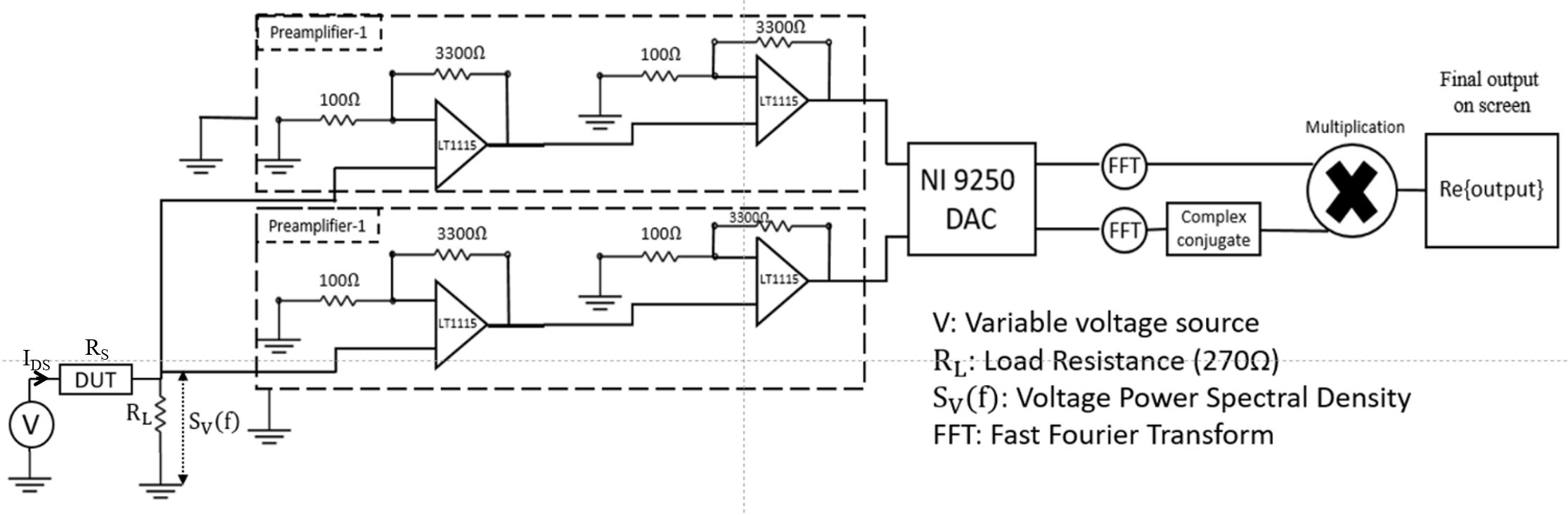}
    \caption{Schematic layout of the low-frequency noise (LFN) cross-correlation spectroscopy data acquisition system. The device under test (DUT) is biased via a variable DC voltage source ($V$) through a series resistance ($R_S$) and load resistor ($R_L = 270\,\Omega$). The noise voltage fluctuations ($S_V(f)$) are simultaneously fed into two parallel, low-noise preamplifier stages (LT1115 op-amp configurations) to eliminate uncorrelated instrumentation noise. The amplified analog signals are digitized by an NI 9250 ADC, processed using Fast Fourier Transforms (FFT) and complex conjugation, and cross-multiplied to compute the real component ($\text{Re}\{\text{output}\}$) of the cross-spectral density.}
	\label{fig:noiseS_2}
\end{figure*}
\begin{figure*}[htbp]
	\centering
	\includegraphics[width=1.0\linewidth]{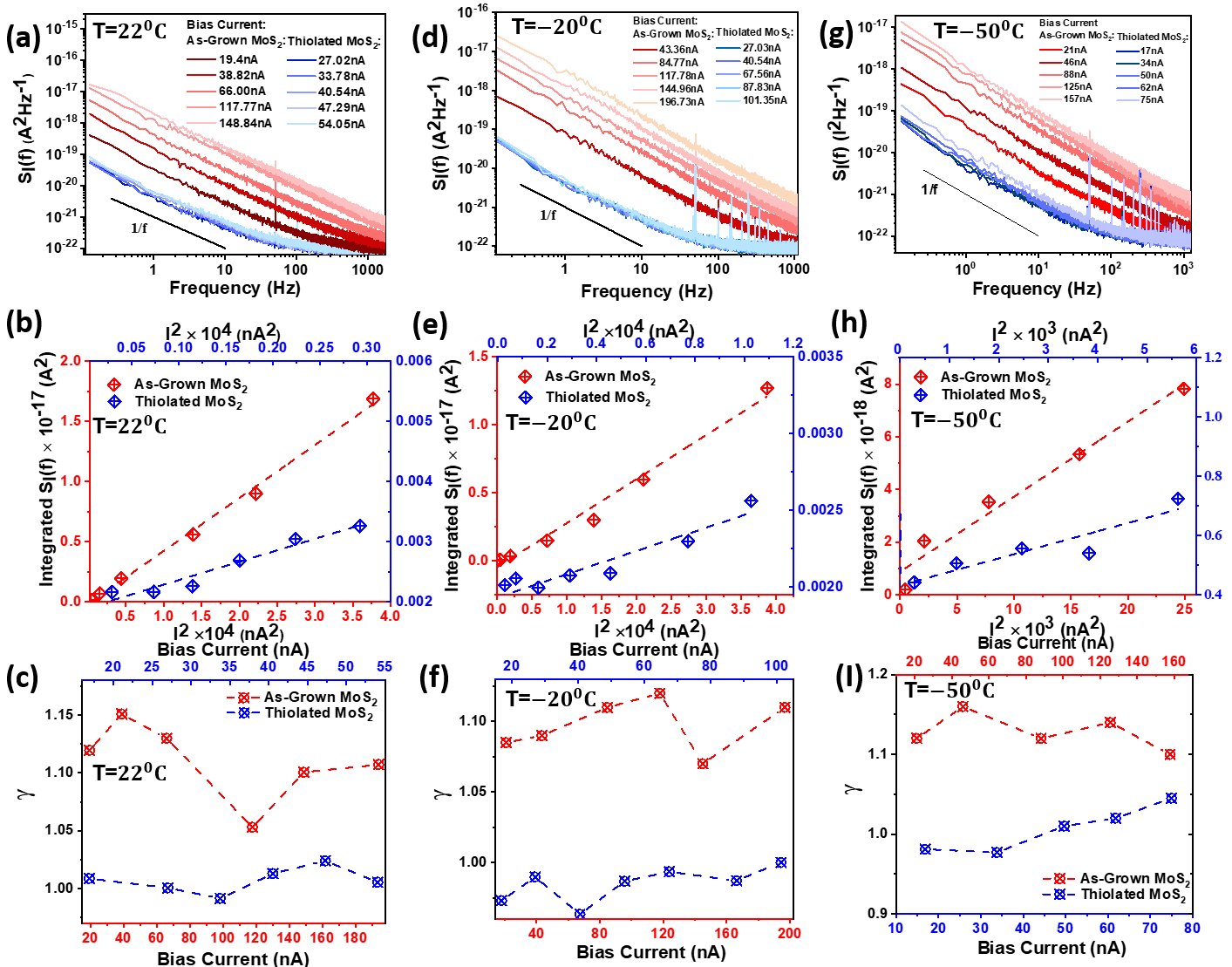}
    \caption{Temperature-dependent low-frequency noise (LFN) characteristics of pristine (as-grown) and thiol-passivated monolayer $\mathrm{MoS}_2$ devices evaluated at $T = 22\,^\circ\text{C}$ (a--c), $T = -20\,^\circ\text{C}$ (d--f), and $T = -50\,^\circ\text{C}$ (g--l). (a,~d,~g)~Current noise power spectral density $S_I(f)$ as a function of frequency across various bias currents, showing the characteristic $1/f$ behavior before (red traces) and after (blue traces) thiol treatment. (b,~e,~h)~Integrated noise power spectral density across the frequency band $0\text{--}10\,\text{Hz}$ as a function of squared channel current ($I^2$) for the as-grown (red diamonds, bottom/left axes) and thiolated (blue diamonds, top/right axes) devices. (c,~f,~l)~Extracted frequency exponent $\gamma$ as a function of bias current for as-grown (red) and thiolated (blue) devices, demonstrating the reduction and stabilization of $\gamma \approx 1.0$ across all temperatures following thiol passivation.}
	\label{fig:noiseS_1}
\end{figure*}
\begin{figure*}
	\centering
	\includegraphics[width=1.0\linewidth]{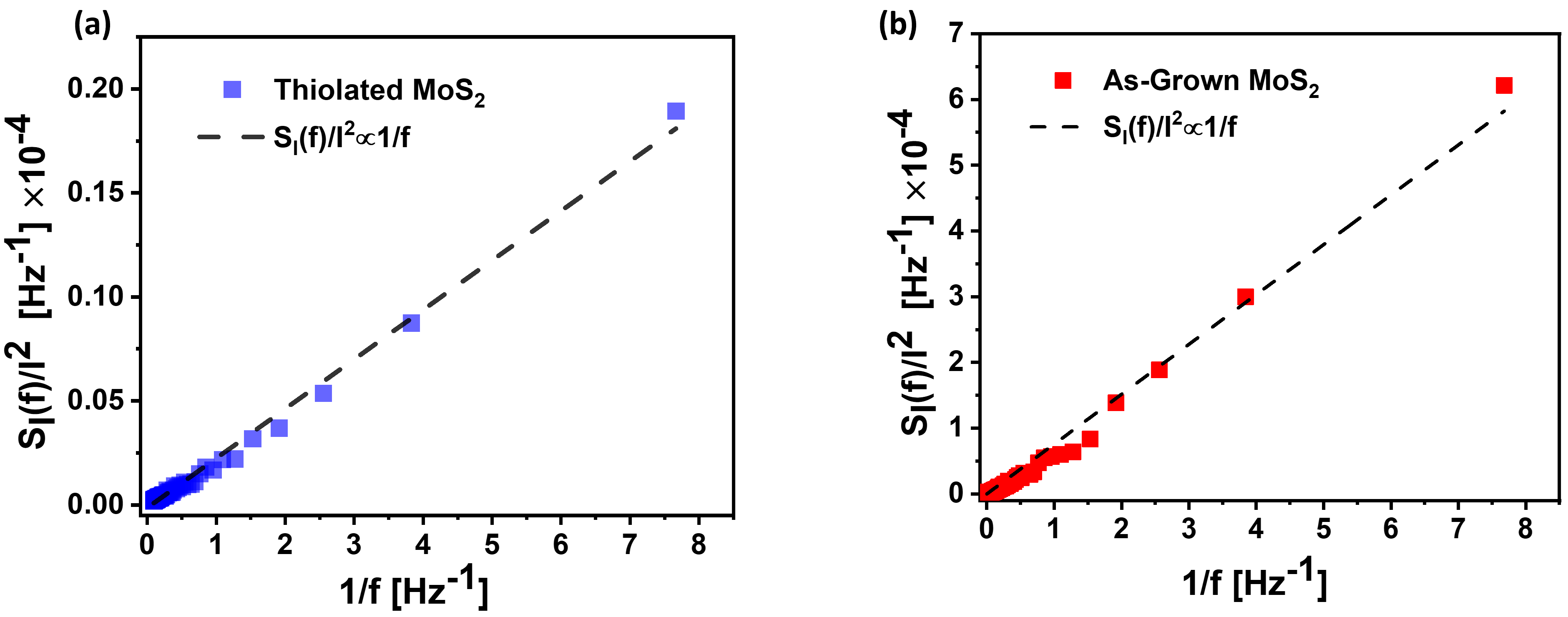}
    \caption{ (a) octane-thiolated and (b) as-grown monolayer MoS$_2$ FETs. The normalized power spectral density, $S_I(f)/I^2$, is plotted as a function of $1/f$, demonstrating the characteristic $1/f$ dependence of the noise. The dashed lines represent linear fits to the experimental data. According to Hooge's relation, $S_I(f)/I^2 = \alpha_H/(Nf)$, the slope of the linear fit corresponds to $\alpha_H/N$, where $\alpha_H$ is Hooge's parameter and $N$ is the total number of charge carriers participating in conduction.}
	\label{fig:linearF}
\end{figure*}

\begin{figure*}
	\centering
	\includegraphics[width=1.0\linewidth]{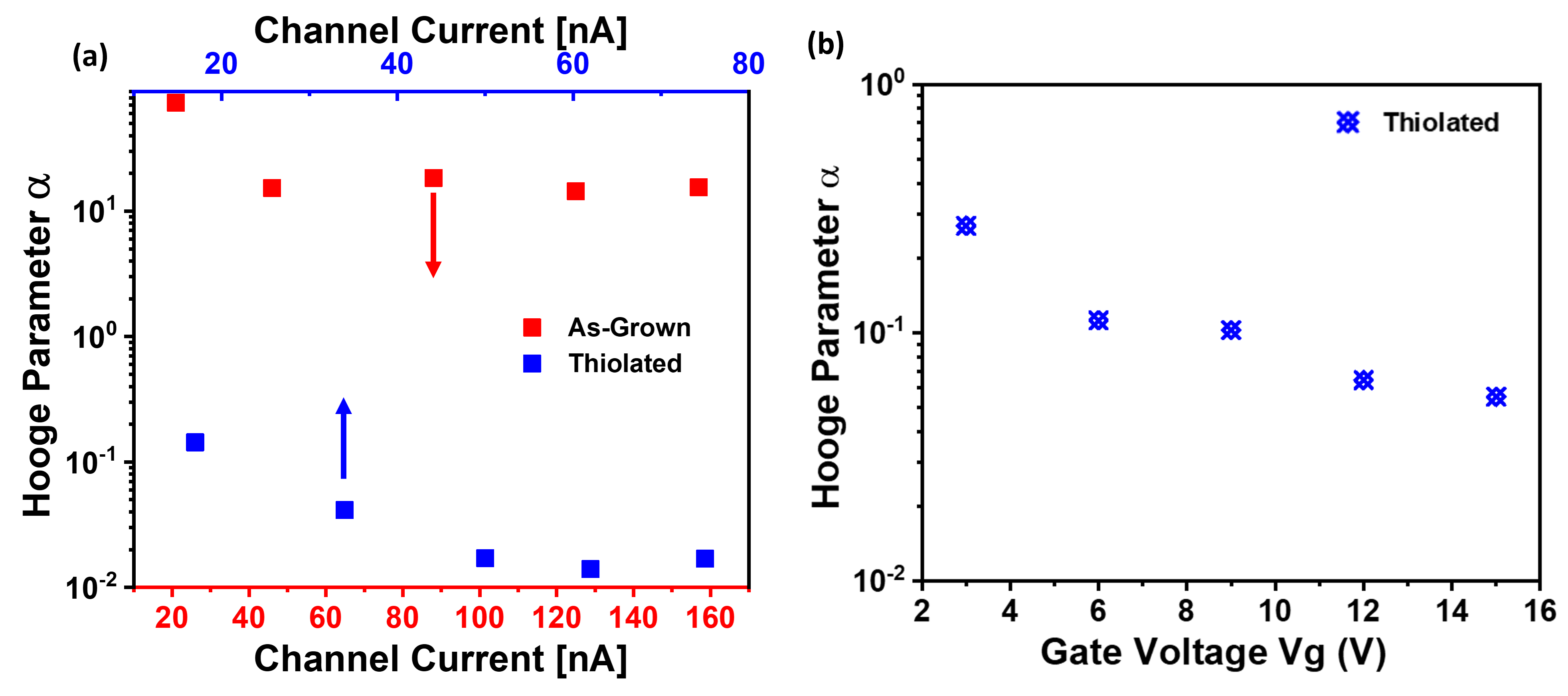}
    \caption{\textbf{Hooge parameter analysis.} 
(a)~Hooge parameter $\alpha$ as a function of channel current for the device in the as-grown state (red squares) and after thiolation (blue squares). 
(b)~Evolution of the Hooge parameter $\alpha$ as a function of back-gate voltage $V_\text{g}$ for the thiolated device.}
	\label{fig:alpha}
\end{figure*}

The device resistance $R_s$ and the load resistance Rl are connected in parallel, as illustrated in the circuit schematic. A voltage source V is applied to drive current through the circuit. The resulting current, $I_{ds}$, represents the channel current flowing through the device, while the corresponding voltage drop across the device is denoted as $V_{ds}$. The voltage drop across the load resistor, $V_{ds}$ , is measured simultaneously. The time-series voltage signal obtained across $R_{L}$ is subsequently amplified using two custom-designed low-noise preamplifiers. These amplified signals are then digitized using an analog-to-digital converter (ADC). The digitized outputs from the two independent channels are Fourier transformed and cross-correlated. The cross-correlated signal yields the power spectral density (PSD) of the parallel combination of the device resistance and the load resistor, denoted as Sv(f). The current power spectral density  Si(f) is then calculated using the appropriate  \cite{Renteria2014},
\begin{equation}
	S_{I}(f)= S_{V}(f)(\frac{R_S+R_L}{R_SR_L})^2
\end{equation} 
while the channel current , Ids is determined from the corresponding equation,
\begin{equation}
	I_{DS}=\frac{V}{R_S+R_L}
\end{equation}
 This cross-correlated noise power spectral density is averaged over 100 data sets to eliminate any random noise present in the circuit. Fig.\ref{fig:noiseS_2} is the schematic diagram used for developing the custom low frequency noise spectroscopy set-up. Fig.\ref{fig:noiseS_1} shows the acquired noise data for devices before and after thiol treatment at different temperatures (22$^0$C, -20$^0$C and -50$^0$C) where (a),(d) and (g) are the current noise power spectral density, (b),(e) and (h) corresponds to the integrated PSD for a frequency 0-10Hz as a function of channel current and (c),(f) and (I) are the calculated freqenecy exponent $\gamma$, both before and after thiol treatment.
 The Hooge parameter ($\alpha_\text{H}$) is a dimensionless quantity that quantifies the intrinsic level of $1/f^\gamma$ (flicker) noise in a conducting medium. The Hooge's empirical relation is given by,
\begin{equation}
    \frac{S_I(f)}{I^2} = \frac{\alpha_\text{H}}{N f^{\gamma}}
    \label{eq:hooge_canonical}
\end{equation}
the normalized current power spectral density ($S_I(f)/I^2$) changes inversely with frequency ($f$) and the total number of free charge carriers ($N$). Within the carrier mobility fluctuation (CMF) framework, higher values of $\alpha_\text{H}$ represents severe momentum relaxation driven by defects and vacancies, whereas lower values signify reduced Coulomb scattering due to effective defect passivation.
In Figure~\ref{fig:linearF}(a,b), $S_I(f)/I_\text{ds}^2$ is plotted as a function of the inverse frequency ($1/f$) for both the thiolated and as-grown MoS$_2$ monolayer devices across varying channel currents. The linear response across the low-frequency band in the 1/f region confirms the canonical $1/f$ scaling predicted by Hooge's emperical relation.  The slope of each trace directly provides the parameter ratio,
\begin{equation}
	\text{Slope} = \frac{\alpha_\text{H}}{N}
\end{equation}
The total number of carriers $N$ participating in conduction was independently determined gated electrical measurements, ($N = nWL$), allowing extraction of $\alpha_\text{H}$. Figure~\ref{fig:alpha}(a) displays the resulting Hooge parameter $\alpha_\text{H}$ plotted against the channel current $I_\text{ds}$. For the unpassivated (as-grown) device, $\alpha_\text{H}$ remains persistently high within the range of $10\text{--}11$ (red squares), which reflects intense Coulomb scattering and mobility fluctuations caused by high concentrations sulphur vacancies and sub-stoichiometric defects. Following 1-octanethiol functionalization, the Hooge parameter drops precipitously down to $\alpha_\text{H} \approx 10^{-2}\text{--}10^{-1}$ (blue squares),an overall attenuation exceeding two orders of magnitude. This reduction in $\alpha_\text{H}$ provides spectroscopic evidence that covalent thiol coordination successfully heals localized vacancies, screening ionized impurity potentials and suppressing long-range Coulombic mobility fluctuations.
Figure~\ref{fig:alpha}(b) illustrates the evolution of $\alpha_\text{H}$ for the thiolated device as a function of back-gate bias ($V_\text{g}$). As the device is biased deeper into the accumulation regime ($V_\text{g} = 3\,\text{V} \rightarrow 15\,\text{V}$), $\alpha_\text{H}$ decreases from $\sim 2.5 \times 10^{-1}$ to $\sim 5 \times 10^{-2}$. This is summarized in Table~\ref{tab:noise_summary}.

\clearpage
\section{S6: Density Functional theory Calculation}
Density Functional Theory (DFT) calculations were performed using the Quantum Espresso software package\cite{Giannozzi2009-jp, Giannozzi2017-vu}. We used an optimised Ultra-Soft Pseudopotentials (USPP) with Scalar-relativistic correction and non-linear core correction for all the atoms and the generalised gradient approximation for the exchange correlation\cite{DalCorso2014}. The monolayer MoS2 was modelled by a supercell, using a 5 × 5 unit cell with the vacuum spacing of 20\text{\AA} along the z-direction (to avoid the interlayer interaction) and to obtain the defect-induced structure, six S atoms are removed to achieve the desired percentage of S defects. To obtain the thiol-treated structure, three 1-octanethiol molecule was places at a one side of the supercell to simulate passivation of S defects at the top-surface of the monolayer. Both of the structures were relaxed using the kinetic energy cutoff for the plane-wave basis was set to 40 Ry, and the convergence threshold for atomic forces was set to $10^{-3}$ eV/atom. Following structural optimization, self-consistent field (SCF) calculations were performed using a 3×3×1 K-point mesh with a convergence threshold of $10^{-6}$ Ry. The Density of States (DoS) was then calculated using a subsequent non-self-consistent field (NSCF) calculation using a 7 x 7 x 3 K-point mesh

\newpage

\bibliography{defectpassivation.bib}
\end{document}